\RequirePackage{ifpdf}
\documentclass[12pt]{JHEP3}
\pdfoutput=1
\usepackage{amsfonts, amsmath, mathtools}
\usepackage{graphicx}
\DeclarePairedDelimiter\bra{\langle}{\rvert}
\DeclarePairedDelimiter\ket{\lvert}{\rangle}
\DeclarePairedDelimiter\bras{[}{\rvert}
\DeclarePairedDelimiter\kets{\lvert}{]}
\DeclarePairedDelimiterX\braket[2]{\langle}{\rangle}{#1 #2}
\DeclarePairedDelimiterX\brakets[2]{[}{]}{#1 #2}
\DeclarePairedDelimiterX\grakets[3]{\langle}{]}{#1 \vert #2 \vert #3}
\DeclarePairedDelimiterX\graket[3]{[}{\rangle}{#1 \vert #2 \vert #3}

\title{The KLT relations in unimodular gravity}
\author{Daniel J. Burger$^{1,2}$\footnote{burgerj.daan@gmail.com}, George F. R. Ellis$^{2}$\footnote{gfrellis@gmail.com}, Jeff Murugan$^{1}$\footnote{jeff.murugan@uct.ac.za}\, \& Amanda Weltman$^{2}$\footnote{awelti@gmail.com}\\
${}^{1}$The Laboratory for Quantum Gravity \& Strings, \\
${}^{2}$The Cosmology \& Gravity Group,\\
Department of Mathematics and Applied Mathematics, \\
University of Cape Town, \\
Private Bag, Rondebosch, 7700, \\
South Africa.}

\abstract{With this article, we initiate a systematic study of some of the symmetry properties of unimodular gravity, building on much of the known structure of general relativity, and utilizing the powerful technology developed in that context. In particular, we show, up to four-points and tree-level, that the KLT relations of perturbative gravity hold for {\it tracefree} or {\it unimodular} gravity.}
 
 \preprint{QGASLAB-15-08}

\begin{document}

\maketitle

\section{Introduction}
One hundred years on, Einstein's General Theory of Relativity (GR) remains {\it the} most remarkable incubator for new ideas in gravitational physics. From cosmology to condensed matter analogues, it has shaped the way we think about our universe across a multitude of scales, in a rather fundamental way. Modulo a handful of important caveats, this understanding has, for the most part, come from studying {\it solutions} to the Einstein field equations.

\noindent 
Arguably, one of the most significant of these exceptional developments of the theory was the realization of GR as an {\it effective field theory} encoding the low energy dynamics of a massless spin-2 particle, the graviton (see, for example, \cite{Deser:2009fq} and references therein). In this language, GR in all its nonlinear glory arises from the central premise of a hard graviton, $h_{\mu\nu}$, propagating on a flat background, $\eta_{\mu\nu}$. At the linearised level, the dynamics of such a spin-2 particle is governed by the Pauli-Fiertz Lagrangian. Self-consistency then forces it to couple to its own energy-momentum tensor if the theory is to admit any coupling to matter. It is precisely this self-interaction that facilitates the bootstrap to the fully nonlinear, nonpolynomial Einstein-Hilbert Lagrangian.

\noindent
Another exception was the discovery that GR, albeit a particular 10-dimensional form of it, arose in the {\it low energy limit} of the critical superstring. In this context, the structure of the Einstein-Hilbert action is determined from the leading order contribution to the vanishing of the beta functions that guarantee the one-loop cancellation of the conformal anomaly of the worldsheet sigma model \cite{Callan:1985ia}. Irrespective of how one feels about superstrings, it is a remarkable feature of the theory that the closed string spectrum contains a massless, spin-2 field with all the characteristics of the graviton, $h_{\mu\nu}$. This was soon followed by the observation that two open strings may be glued together to form a closed string \cite{Kawai:1985xq}. This seemingly innocuous result that an
\begin{eqnarray}
  |\mathrm{closed\,\,string\,\,state}\rangle = |\mathrm{open\,\,string\,\,state}\rangle
  \otimes |\mathrm{open\,\,string\,\,state}\rangle\nonumber
\end{eqnarray}
has precipitated a silent revolution in our understanding of the nature of gravity that can be summarised schematically as 
\begin{eqnarray}
  \mathrm{gravity}\,\sim\,(\mathrm{gauge\,\,theory})^{2},\nonumber
\end{eqnarray}
and whose precise statement is embodied in the KLT relations that connect gauge theory and gravity at the level of {\it scattering amplitudes} (see, for example, \cite{Bern:2002kj} for an outstanding review of the state of the art). For 4, 5 and 6-point tree-level scattering, for example, these KLT equations read
\cite{Elvang:2013cua}
\begin{eqnarray}
  M_{4}(1234) &=& -s_{12}A_{4}[1234]A_{4}[1243],\nonumber\\
  M_{5}(12345) &=& s_{23}s_{45}A_{5}[12345]A_{5}[13254] + (3\leftrightarrow 4),\\
  M_{6}(123456) &=& -s_{12}s_{45}A_{6}[123456]\left(s_{35}A_{6}[153462] + (s_{34}+s_{35})A_{6}[154362]\right) 
  + \mathcal{P}(2,3,4),\nonumber
  \label{KLT}
\end{eqnarray}
where gravity amplitudes are denoted $M_{n}$, gauge theory amplitudes are $A_{n}$ and $\mathcal{P}$ denotes a permutation of the legs in the scattering diagram. Importantly, while this connection was first realised in, it is by no means restricted to string theory. Indeed, since their discovery, the KLT relations have revealed similar connections between 4-dimensional GR and Yang-Mills theory; 4-dimensional axion-dilaton gravity and Yang-Mills theory and even $\mathcal{N}=8$ supergravity and $\mathcal{N}=4$ super Yang-Mills theory \cite{Elvang:2013cua}, where it has proven particularly useful in probing the UV finiteness of the supergravity theory. Before elaborating on the focal point of this article, it is worth noting that this story features two crucial supporting cast members; the {\it spinor helicity formalism} and {\it twistor calculus} that provide the essential mathematical tools and render the computation of scattering amplitudes on both sides of the KLT map tractable \cite{Witten:2003nn}. Suffice it to say, the KLT relations and their generalisations have led to a completely novel way of looking at gravity at both the quantum and classical levels that call into question our understanding of such foundational ideas such as locality, causality and perhaps even spacetime itself \cite{Arkani-Hamed:2013jha}. The goal of this article, however, is far less lofty; we look only to answer the question:
\begin{center}
{\it Do the KLT relations still hold for modifications of GR?}
\end{center}
Of course, contemporary cosmology is replete with theories that fall into the category of ``modified gravity": $f(R), f(G), f(T), f(\#\mathrm{yourfavouritescalarinvariant})$, massive gravity, new massive gravity, Lovelock gravity, pure Lovelock gravity and braneworld gravity, to name but a few. We will focus on one particular modification, {\it tracefree} or {\it unimodular} gravity, (UG) that goes all the way back to Einstein in 1919 but which was resurrected most recently in \cite{Weinberg:1988cp}, in the context of the cosmological constant problem. While it does not resolve the issue of the cosmological constant, UG does relegate it to an integration constant to be fixed by empirical data. It does so by decoupling fluctuations of the quantum vacuum from gravitational physics rendering an entirely viable {\it classical} theory of gravity \cite{Ellis:2010uc}. In fact, at the classical level, UG is expected to be completely indistinguishable from GR \cite{Alvarez:2012uz,Alvarez:2015pla} (see also the extended discussion in \cite{Padilla:2014yea}) even though the former only preserves a Weyl transverse subgroup, {\it WTDiff(M)} of the full {\it Diff(M)} symmetry group of GR. This difference will, however, manifest at the quantum level. With an ultimate goal of exploring the quantum differences between unimodular gravity and GR in mind, it is certainly important to understand the extent of their similarities. Toward this end, we refine our question above to 
\begin{center}
{\it To what extent do the KLT relations hold in unimodular gravity?}
\end{center}
The rest of this article is structured as follows. In the interests of pedagogy, the next section is devoted to a detailed derivation of the perturbative structure of both the Einstein-Hilbert and unimodular Lagrangians, paying particular attention to the symmetries encoded in each. 
Section 3 contains the main body of our argument, namely the computation of graviton scattering amplitudes in unimodular gravity. These are compared to the corresponding results in GR and used to show that, up to four points, the KLT relations are satisfied in a nontrivial way in unimodular gravity. We conclude in section 4. with a discussion of the result and some speculations on generalisation to higher loops and the inclusion of matter. Finally, and again in the interests of presenting our work in as pedagogical a fashion as possible, we give a concise introduction to the spinor helicity formalism for a gravity readership in the appendix.

\section{The perturbative structure of GR and UG}
\renewcommand{\theequation}{\thesection.\arabic{equation}}
\subsection{GR Lagrangian}
We include this section purely in the interests being self contained, and to establish our notation. Readers familiar with the weak field expansion of the Einstein-Hilbert action are invited to skip ahead. While the treatment given here can certainly be extended to generally curved backgrounds, we will restrict our attention to a graviton propagating on a flat Minkowski geometry whose metric tensor $\eta_{\mu\nu} = \mathrm{diag}(-1,+1,+1,+1)$. Consequently, we take $g_{\mu\nu} = \eta_{\mu\nu} + \kappa h_{\mu\nu}$ and expand the Einstein-Hilbert action
\begin{align} \label{Lagrangian_EH}
  S_{EH} = \frac{1}{2\kappa^{2}}\int d^{4}x\,\sqrt{|g|}\, R,
\end{align}
in powers of $\kappa h_{\mu\nu}$. Here, as is standard in the literature, we define the gravitational coupling as $\kappa^{2}\equiv 8\pi G_{N}$. Unlike, for example, Maxwell electrodynamics, this expansion generates an infinite series in $h_{\mu\nu}$, due essentially to the presence of the inverse metric in the Ricci scalar 
\begin{eqnarray}
  R = g^{\mu\nu} \left( \partial_{\nu}\Gamma^{\lambda}_{\mu\lambda} - \partial_{\lambda}\Gamma^{\lambda}_{\mu\nu} +  
  \Gamma^{\tau}_{\mu\lambda}\Gamma^{\lambda}_{\tau\nu} - \Gamma^{\tau}_{\mu\nu}\Gamma^{\lambda}_{\tau\lambda}   
  \right),
  \label{Ricci_Scalar}
\end{eqnarray}
and the square root of the determinant of the metric in the volume form. Expanded in $h_{\mu\nu}$ up to cubic order, these
factors contribute
\begin{eqnarray}
  g^{\mu\nu} = \eta^{\mu\nu} - \kappa h^{\mu\nu} + \kappa^{2} h^{\mu\lambda}h_{\lambda}^{\nu} -
  \kappa^{3} h^{\mu\lambda}h_{\lambda\rho}h^{\rho\nu} + O(h^{4}),
\end{eqnarray} 
and, respectively,
\begin{eqnarray}
  \sqrt{|g|} &=& \prod^{\infty}_{n=1} \left[ \sum^{\infty}_{m=0} \left( \frac{1}{m!} \left[ \frac{(-1)^{n-1}}{2n} \left( h^n \right)^{\mu}_{\mu} \right]  \right) \right] \\
  &=&  1 + \frac{\kappa}{2} h + \frac{\kappa^{2}}{8} \left( h^2 -2h^{\mu\nu}h_{\mu\nu}  \right) 
  + \frac{\kappa^{3}}{48} \left( h^3 -6h h^{\mu\nu}h_{\mu\nu} +8 h^{\mu\nu}h_{\nu\lambda}h^{\lambda\mu} \right) 
  + O(h^4),\nonumber
\end{eqnarray} 
Substituting this into the Einstein-Hilbert Lagrangian and organizing the resulting expansion in powers of 
$h_{\mu\nu}$ gives the formal series,
\begin{eqnarray}
  \mathcal{L} = \mathcal{L}_2 + \kappa \mathcal{L}_3 + \kappa^{2} \mathcal{L}_4 + ...
\end{eqnarray}
In order to extract the Feynman rules from this Lagrangian, we still need to fix a gauge. A common choice in the perturbative gravity literature is the {\it de Donder gauge} in which $\partial_{\mu}h^{\mu\nu} = \frac{1}{2}\partial^{\nu}
h_{\mu}^{\mu}$ bringing the quadratic terms in the Lagrangian into the form $-\frac{1}{2}h_{\mu\nu}\,\square h^{\mu\nu} + \frac{1}{4}h_{\mu}^{\,\,\mu}\,\square h_{\nu}^{\,\,\nu}$. This will facilitate reading off of the de Donder gauge propagator a little later but at this point, up to quartic order (good for 4-point graviton scattering), the perturbative Lagrangian of GR in all its indicial glory, reads
\begin{align} \label{Lagrangian_Perturbative_GR_dD}
\begin{split}
\mathcal{L}_2
&=
- \frac{1}{8}\partial_{\alpha} h\, \partial^{\alpha} h 
+ \frac{1}{4} \partial_{\gamma} h^{\alpha\beta}\,\partial^{\gamma} h_{\alpha\beta} \\
\mathcal{L}_3
&=
+ \frac{1}{4} h^{\alpha\gamma} \,\partial_{\beta} h \,\partial^{\beta} h_{\alpha\gamma}
- \frac{1}{16} h\, \partial_{\beta} h \,\partial^{\beta} h
- \frac{1}{4} h^{\alpha\gamma} \,\partial_{\alpha} h_{\beta\delta} \,\partial_{\gamma} h^{\beta\delta}\\
&
+ \frac{1}{2} h^{\alpha\gamma}\, \partial_{\gamma} h_{\beta\delta}\, \partial^{\delta} h_{\alpha}^{\,\,\beta}
- \frac{1}{2} h^{\alpha\gamma} \,\partial_{\delta} h_{\gamma\beta}\, \partial^{\delta} h_{\alpha}^{\,\,\beta}
+ \frac{1}{8} h\, \partial^{\delta} h_{\alpha\beta}\, \partial_{\delta} h^{\alpha\beta}\\
\mathcal{L}_4
&=
  \frac{1}{4} h^{\alpha\delta}h^{\beta\mu}\,\partial_{\nu} h_{\delta\mu}\,\partial^{\nu} h_{\alpha\beta}
- \frac{1}{4} h^{\alpha\delta}h^{\beta\mu}\,\partial_{\nu} h_{\beta\mu}\,\partial^{\nu} h_{\alpha\delta}
- \frac{1}{16} h^{\alpha\delta}h^{\alpha\delta}\, \partial_{\nu} h_{\beta\mu}\, \partial^{\nu} h_{\beta\mu}
+ \frac{1}{32} h h\, \partial_{\nu} h_{\alpha\beta}\, \partial^{\nu} h^{\alpha\beta}\\
&
- \frac{1}{2} h_{\alpha}^{\beta} h^{\alpha\delta}\,\partial_{\beta} h_{\mu\nu}\,\partial^{\nu} h_{\delta}^{\mu}
+ \frac{1}{2} h h^{\delta\beta} \,\partial_{\beta} h_{\mu\nu} \,\partial^{\nu} h_{\delta}^{\mu}
+ \frac{1}{2} h_{\alpha}^{\beta}h^{\alpha\delta}\,\partial_{\nu} h_{\beta\mu}\,\partial^{\nu} h_{\delta}^{\mu}
- \frac{1}{4} h h^{\delta\beta} \,\partial_{\nu} h_{\beta\mu}\,\partial^{\nu} h_{\delta}^{\mu}\\
&
- \frac{1}{2} h^{\alpha\delta}h^{\beta\mu}\,\partial_{\beta} h_{\alpha}^{\nu}\,\partial_{\delta} h_{\mu\nu}
+ \frac{1}{4} h_{\alpha}^{\beta} h^{\alpha\delta}\,\partial_{\beta} h_{\mu\nu}\,\partial_{\delta} h^{\mu\nu}
- \frac{1}{8} h h^{\delta\beta}\,\partial_{\beta} h_{\mu\nu}\,\partial_{\delta} h^{\mu\nu}
- \frac{1}{8} h_{\alpha}^{\beta} h^{\alpha\delta}\,\partial_{\beta} h\,\partial_{\delta} h\\
&
+ \frac{1}{2} h^{\alpha\delta} h^{\beta\mu}\,\partial^{\nu} h_{\alpha\delta}\,\partial_{\mu} h_{\beta\nu}
+ \frac{1}{8} h_{\alpha\delta} h^{\alpha\delta}\,\partial^{\nu} h^{\beta\mu}\,\partial_{\mu} h_{\beta\nu}
- \frac{1}{16} h h \,\partial^{\nu} h^{\beta\mu}\,\partial_{\mu} h_{\beta\nu}
+ \frac{1}{4} h h^{\delta\beta}\partial^{\nu} h_{\delta}^{\mu}\partial_{\mu} h_{\beta\nu}\\
&
- \frac{1}{4} h^{\alpha\delta} h^{\beta\mu}\,\partial_{\beta} h_{\alpha\delta}\,\partial_{\mu} h
+ \frac{1}{4} h_{\alpha}^{\beta} h^{\alpha\delta}\,\partial_{\beta} h_{\delta}^{\mu}\,\partial_{\mu} h
- \frac{1}{4} h h^{\delta\beta} \,\partial_{\beta} h_{\delta}^{\mu}\,\partial_{\mu} h
+ \frac{1}{4} h^{\alpha\delta} h^{\beta\mu}\, \partial_{\delta} h_{\alpha\beta}\,\partial_{\mu} h\\
&
+ \frac{1}{2} h^{\alpha\delta} h^{\beta\mu} \,\partial_{\beta} h_{\alpha}^{\nu}\,\partial_{\mu} h_{\delta\nu}
- \frac{1}{2} h^{\alpha\delta} h^{\beta\mu} \,\partial^{\nu} h_{\alpha\beta}\,\partial_{\mu} h_{\delta\nu}
- \frac{1}{4} h_{\alpha}^{\beta} h^{\alpha\delta} \,\partial_{\mu} h\, \partial^{\mu} h_{\delta\beta}
+ \frac{1}{8} h h^{\delta\beta}\, \partial_{\mu} h\, \partial^{\mu} h_{\delta\beta},
\end{split}
\end{align}
and will form the basis for comparison to unimodular gravity below.
\subsection{Unimodular Lagrangian}
Before writing down the equivalent perturbative expansion for unimodular gravity, it will be useful to recall the 
symmetries of the theory. The Einstein-Hilbert of GR is famously invariant under the full group of diffeomorphisms on the spacetime manifold, Diff($M$), under which 
$g_{\mu\nu} \rightarrow g_{\mu\nu} + \nabla_{\mu}\xi_{\nu} + \nabla_{\nu}\xi_{\mu}$ or, in infinitesimal form, $h_{\mu\nu} \rightarrow h_{\mu\nu} + \partial_{\mu}\xi_{\nu} + \partial_{\nu}\xi_{\mu}$. The defining characteristic of Einstein's 1919 tracefree theory is that the metric determinant is held fixed, to unity in the special case of unimodular gravity. This unimodularity condition breaks Diff($M$) to the proper subgroup of {\it transverse} diffeomorphisms, TDiff($M$) under which $h_{\mu\nu} \rightarrow h_{\mu\nu} + \partial_{\mu}\xi_{\nu} + \partial_{\nu}\xi_{\mu}$, with $\partial_{\mu}\xi^{\mu} =0$. This is, of course, just a classical gauge fixing of the GR action and the reason why the two theories are classically indistinguishable\footnote{The reader will no doubt have noticed that we have also not been distinguishing between ``tracefree" and ``unimodular" since, from our perspective, the only difference between the two is the value of the constant that the determinant of the metric is fixed at. This will have no effect on any scattering amplitudes.} \cite{Ellis:2010uc,Padilla:2014yea} (modulo the important issue of the interpretation of the cosmological constant). In fact the theory enjoys an additional Weyl symmetry under which $g_{\mu\nu}\to e^{2\sigma(x)}g_{\mu\nu}$, enhancing its symmetry to WTDiff($M$) with a corresponding four generators per spacetime point.

There are, in fact, many ways of implementing the WTDiff($M$) symmetry into an action functional that range from the most obvious enforcing of the fixed metric determinant through a Lagrange multiplier \cite{LopezVillarejo:2010iq} so that 
\begin{eqnarray}
  S_{EH}\to S_{UG} = \int_{M}d^{4}x\,\,\left[\sqrt{|g|}R + \lambda\left(\sqrt{|g|}-1\right) \right],
\end{eqnarray}
to Henneaux and Teitelboim's \cite{Henneaux:1989zc} more sophisticated formulation in which the tracefree equations are derived from the fully covariant action
\begin{eqnarray}
  S_{HT} = \frac{1}{2\kappa^{2}}\left(\int_{M} d^{4}x\,\,\sqrt{|g|}(R+2\lambda) + \int_{M}A_{3}\wedge d\lambda\right),
\end{eqnarray}
where $A_{3}$ and $\lambda$ are a spacetime 3-form and scalar respectively. All these formalisms have been treated extensively in the literature, and all produce Einstein's 1919 equations\footnote{Historically, this is not entirely correct since Einstein's actual 1919 equations did not take into account that the energy momentum tensor on the right hand side of this equation must also be tracefree.}
\begin{eqnarray}
  R_{\mu\nu} - \frac{1}{4}g_{\mu\nu}R = 8\pi G_{N} \widetilde{T}_{\mu\nu},
\end{eqnarray}
with $\mathrm{tr}\left(\widetilde{T}_{\mu\nu}\right)=0$. Since we are interested making {\it on-shell} statements about the theory, any of these various action principles will suffice for our purposes. However, for convenience, we will use the one that begins with a rescaling of $g_{\mu\nu}\to \hat{g}_{\mu\nu}\equiv g^{-1/4}g_{\mu\nu}$. The resulting action, formulated in terms of $\hat{g}_{\mu\nu}$ is unimodular since $\hat{g} = \mathrm{det}(\hat{g}_{\mu\nu}) = 1$ and reads quite simply,
\begin{eqnarray}
  S_{UG} = \frac{1}{2\kappa^{2}}\int_{M}d^{4}x\,\, \hat{R}(\hat{g}_{\mu\nu}).
  \label{unimodular-action}
\end{eqnarray}
The perturbative expansion for unimodular gravity then proceeds in much the same way from (\ref{unimodular-action}) as 
that for GR follows from the Einstein-Hilbert action. Again, we write $\hat{g}_{\mu\nu} = \eta_{\mu\nu} + \hat{h}_{\mu\nu}$ except that now $\hat{h}_{\mu\nu} = h_{\mu\nu}-\frac{1}{4}h\eta_{\mu\nu}$ is traceless. This cannot be interpreted as a field redefinition since the trace, $h$, of $h_{\mu\nu}$ cannot be recovered from $\hat{h}$. This substitution then yields the perturbative Lagrangian (by order in $h$) for unimodular gravity. Formally, $\hat{\mathcal{L}}=\hat{\mathcal{L}}_{2}+\kappa \hat{\mathcal{L}}_{3}+ ...$, where for example,
\begin{eqnarray} 
\label{Lagrangian_Perturbative_TFG}
\hat{\mathcal{L}}_2
&=&
- \frac{3}{32}\partial_{\alpha} h \partial^{\alpha} h 
+ \frac{1}{4} \partial_{\gamma} h^{\alpha\beta}\partial^{\gamma} h_{\alpha\beta}
+ \frac{1}{4} \partial^{\alpha} h \partial^{\beta} h_{\alpha\beta}
- \frac{1}{2} \partial_{\alpha} h^{\alpha\beta}\partial^{\gamma} h_{\gamma\beta}\nonumber \\
\hat{\mathcal{L}}_3
&=&
+ \frac{3}{8} h^{\alpha\gamma} \partial_{\beta} h \partial^{\beta} h_{\alpha\gamma}
- \frac{17}{128} h \partial_{\beta} h \partial^{\beta} h
- \frac{1}{4} h^{\alpha\gamma} \partial^{\beta} h \partial^{\gamma} h_{\alpha\beta}
- \frac{1}{4} h^{\alpha\gamma} \partial_{\alpha} h^{\beta\delta} \partial_{\gamma} h_{\beta\delta}
+ \frac{5}{32} h^{\alpha\gamma} \partial_{\alpha} h \partial_{\gamma} h \nonumber\\
&{}&
- \frac{1}{2} h^{\alpha\gamma} \partial^{\beta} h_{\alpha\beta} \partial_{\gamma} h
- \frac{1}{2} h^{\alpha\gamma} \partial^{\beta} h_{\alpha\gamma} \partial^{\delta} h_{\beta\delta}
+ \frac{5}{16} h \partial^{\beta} h \partial^{\alpha} h_{\alpha\beta}
+ \frac{1}{2} h^{\alpha\gamma} \partial_{\gamma} h_{\alpha\beta} \partial_{\delta} h^{\beta\delta}
- \frac{1}{4} h \partial_{\gamma} h^{\gamma\beta} \partial^{\delta} h_{\beta\delta}\nonumber\\
&{}&
+ \frac{1}{2} h^{\alpha\gamma} \partial^{\beta} h_{\alpha\beta} \partial^{\delta} h_{\gamma\delta}
+ \frac{1}{2} h^{\alpha\gamma} \partial_{\gamma} h_{\beta\delta} \partial^{\delta} h_{\alpha}^{\beta}
- \frac{1}{2} h^{\alpha\gamma} \partial_{\delta} h_{\gamma\beta} \partial^{\delta} h_{\alpha}^{\beta}
- \frac{1}{8} h \partial_{\beta} h_{\gamma\delta} \partial^{\delta} h^{\gamma\beta}
+ \frac{3}{16} h \partial_{\delta} h_{\gamma\beta} \partial^{\delta} h^{\gamma\beta}.\nonumber
\end{eqnarray}
As a check that we do indeed have the correct invariance required of the theory, let's consider the quadratic piece $\hat{\mathcal{L}}_2$, from which we will derive the propagator. Under a general field redefinition $h_{\mu\nu} \rightarrow h_{\mu\nu} + \delta h_{\mu\nu}$, and up to total derivatives, 
\begin{eqnarray}
  \delta\hat{\mathcal{L}}_2 =
  + \frac{3}{16} \delta h \partial_{\alpha}\partial^{\alpha} h 
  - \frac{1}{2} \delta h^{\alpha\beta} \partial_{\gamma}\partial^{\gamma} h_{\alpha\beta}
  - \frac{1}{4} \left( \delta h \partial^{\alpha}\partial^{\beta} h_{\alpha\beta} +\delta h_{\alpha\beta}\partial^{\alpha}  
  \partial^{\beta} h \right)
  + \frac{1}{2} \delta h^{\alpha\beta}\partial^{\gamma}\partial_{(\alpha} h_{\beta)\gamma}.\nonumber
\end{eqnarray}
Evidently, under the restricted set of gauge transformations $\delta h_{\alpha\beta} \rightarrow 2\partial_{(\alpha}\xi_{\beta)} + \frac{1}{2} \eta_{\mu\nu} \phi$ with the parameters obeying the transversality condition $\partial_{\alpha}\xi^{\alpha}=0$, the first and third terms as well as a combination of the second and fourth terms are all invariant. As promised, the traceless perturbative Lagrangian is WTDiff-invariant. We have checked that $\hat{\mathcal{L}}_3$ (and higher order in $h$ terms) also exhibit the same invariance under WTDiff($M$). It remains only to fix the additional gauge redundancies by applying de Donder gauge again, yielding
\begin{eqnarray} 
  \label{Lagrangian_Perturbative_TFG_dD}
   \hat{\mathcal{L}}_2
   &=&
   - \frac{3}{32}\partial_{\alpha} h \partial^{\alpha} h 
   + \frac{1}{4} \partial_{\gamma} h^{\alpha\beta}\partial^{\gamma} h_{\alpha\beta}\nonumber \\
   \hat{\mathcal{L}}_3
   &=&
   + \frac{1}{8} h^{\alpha\gamma} \partial_{\beta} h \partial^{\beta} h_{\alpha\gamma}
   - \frac{5}{128} h \partial_{\beta} h \partial^{\beta} h
   - \frac{1}{4} h^{\alpha\gamma} \partial_{\alpha} h_{\beta\delta} \partial_{\gamma} h^{\beta\delta}
   + \frac{1}{2} h^{\alpha\gamma} \partial_{\gamma} h_{\beta\delta} \partial^{\delta} h_{\alpha}^{\beta}\nonumber\\
   &{}&
   - \frac{1}{2} h^{\alpha\gamma} \partial_{\delta} h_{\gamma\beta} \partial^{\delta} h_{\alpha}^{\beta} 
   + \frac{3}{16} h \partial^{\delta} h_{\alpha\beta} \partial_{\delta} h^{\alpha\beta}
   - \frac{1}{8} h \partial_{\beta} h^{\alpha\gamma} \partial_{\gamma} h_{\alpha}^{\beta}
   + \frac{1}{32} h^{\alpha\gamma} \partial_{\alpha} h \partial_{\gamma} h \\
   \hat{\mathcal{L}}_4
   &=&
   \frac{1}{4} h^{\alpha\delta}h^{\beta\mu}\partial_{\nu} h_{\delta\mu}\partial^{\nu} h_{\alpha\beta}
   - \frac{1}{4} h^{\alpha\delta}h^{\beta\mu}\partial_{\nu} h_{\beta\mu}\partial^{\nu} h_{\alpha\delta}
   - \frac{1}{16} h^{\alpha\delta}h^{\alpha\delta} \partial_{\nu} h_{\beta\mu} \partial^{\nu} h_{\beta\mu}
   + \frac{7}{64} h h \partial_{\nu} h_{\alpha\beta} \partial^{\nu} h^{\alpha\beta}\nonumber\\
   &{}&
   - \frac{1}{2} h_{\alpha}^{\beta} h^{\alpha\delta}\partial_{\beta} h_{\mu\nu}\partial^{\nu} h_{\delta}^{\mu}
   + \frac{5}{8} h h^{\delta\beta} \partial_{\beta} h_{\mu\nu} \partial^{\nu} h_{\delta}^{\mu}
   + \frac{1}{2} h_{\alpha}^{\beta}h^{\alpha\delta}\partial_{\nu} h_{\beta\mu}\partial^{\nu} h_{\delta}^{\mu}
   - \frac{1}{2} h h^{\delta\beta} \partial_{\nu} h_{\beta\mu}\partial^{\nu} h_{\delta}^{\mu}\nonumber\\
   &{}&
   - \frac{1}{2} h^{\alpha\delta}h^{\beta\mu}\partial_{\beta} h_{\alpha}^{\nu}\partial_{\delta} h_{\mu\nu}
   + \frac{1}{4} h_{\alpha}^{\beta} h^{\alpha\delta}\partial_{\beta} h_{\mu\nu}\partial_{\delta} h^{\mu\nu}
   - \frac{1}{4} h h^{\delta\beta}\partial_{\beta} h_{\mu\nu}\partial_{\delta} h^{\mu\nu}
   - \frac{1}{16} h_{\alpha}^{\beta} h^{\alpha\delta}\partial_{\beta} h\partial_{\delta} h\nonumber\\
   &{}&
   + \frac{1}{2} h^{\alpha\delta} h^{\beta\mu}\partial^{\nu} h_{\alpha\delta}\partial_{\mu} h_{\beta\nu}
   + \frac{1}{8} h_{\alpha\delta} h^{\alpha\delta}\partial^{\nu} h^{\beta\mu}\partial_{\mu} h_{\beta\nu}
   - \frac{1}{8} h h \partial^{\nu} h^{\beta\mu}\partial_{\mu} h_{\beta\nu}
+ \frac{1}{8} h h^{\delta\beta}\partial^{\nu} h_{\delta}^{\mu}\partial_{\mu} h_{\beta\nu}\nonumber\\
&{}&
- \frac{1}{4} h^{\alpha\delta} h^{\beta\mu}\partial_{\beta} h_{\alpha\delta}\partial_{\mu} h
+ \frac{1}{8} h_{\alpha}^{\beta} h^{\alpha\delta}\partial_{\beta} h_{\delta}^{\mu}\partial_{\mu} h
- \frac{1}{4} h h^{\delta\beta} \partial_{\beta} h_{\delta}^{\mu}\partial_{\mu} h
+ \frac{1}{4} h^{\alpha\delta} h^{\beta\mu} \partial_{\delta} h_{\alpha\beta}\partial_{\mu} h\nonumber\\
&{}&
+ \frac{1}{2} h^{\alpha\delta} h^{\beta\mu} \partial_{\beta} h_{\alpha}^{\nu}\partial_{\mu} h_{\delta\nu}
- \frac{1}{2} h^{\alpha\delta} h^{\beta\mu} \partial^{\nu} h_{\alpha\beta}\partial_{\mu} h_{\delta\nu}
- \frac{1}{4} h_{\alpha}^{\beta} h^{\alpha\delta} \partial_{\mu} h \partial^{\mu} h_{\delta\beta}
+ \frac{1}{4} h h^{\delta\beta} \partial_{\mu} h \partial^{\mu} h_{\delta\beta}\nonumber\\
&{}&
+ \frac{1}{16} h h^{\delta\beta} \partial_{\beta} h \partial_{\delta} h
+ \frac{1}{128} h^{\alpha\delta} h_{\alpha\delta} \partial^{\mu} h \partial_{\mu} h
- \frac{13}{512} h h \partial_{\mu} h \partial^{\mu} h.\nonumber
\end{eqnarray}
At this point it is worth noticing that, on comparison with the corresponding expression for the gauge-fixed form for GR,  the Lagrangians differ only by numerical coefficients in terms involving $h$; the index structure of the terms in the overall expression remain unchanged. This has important bearing on what follows.

\section{Amplitudes}
\renewcommand{\theequation}{\thesection.\arabic{equation}}
\subsection{Propagators} \label{Propagators}
Before deriving expressions for the vertices for graviton scattering central to the computation of amplitudes in the theory, we first need to find the appropriate expressions for the graviton propagator which itself derives from the quadratic contribution to the perturbative Lagrangian. The quadratic terms in GR and UG differ only in the coefficient of the term containing factors of the trace $h = h^{\mu}_{\mu}$, so we expect that the computation of the propagator itself will be nearly identical. We will content ourselves with deriving this in GR, and then deducing the corresponding expression in UG. To this end, consider the gauge fixed expression for $\mathcal{L}_2$ from \eqref{Lagrangian_Perturbative_GR_dD} which is of the form,
\begin{eqnarray}
  \mathcal{L}_{2} &=& \frac{1}{2} \partial_{\gamma} h_{\alpha\beta}\, V^{\alpha\beta\mu\nu}\, 
  \partial^{\gamma} h_{\mu\nu},
\end{eqnarray}
with $\displaystyle V^{\alpha\beta\mu\nu} \equiv \frac{1}{4} \eta^{\alpha\beta}\eta^{\mu\nu} - \frac{1}{2} \eta^{\alpha\mu}\eta^{\beta\nu}$. Recognising that the the right hand side of this expression is symmetric with respect to $\alpha\leftrightarrow\beta$, $\mu\leftrightarrow\nu$ and $(\alpha\beta)\leftrightarrow(\mu\nu)$ allows us to trade the rank-2
tensor $h_{\mu\nu}$ for a vector $\Psi_{i}$ where, since there are only ten independent combinations of $\alpha\beta$ and $\mu\nu$ we use the following translation between tensor and vector indices
\begin{eqnarray}
\label{Propagator_Key}
\begin{tabular}{c|c|c|c|c|c|c|c|c|c|c} 
$\alpha\beta,\mu\nu$ & 00 & 11 & 22 & 33 & 01 & 02 & 03 & 12 & 13 & 23\\
\hline
i,j & 1 & 2 & 3 & 4 & 5 & 6 & 7 & 8 & 9 & 10 
\end{tabular}
\end{eqnarray}
With this dictionary in place, the quadratic Lagrangian reads
\begin{eqnarray}
\mathcal{L}_{2} &=& \frac{1}{2} \sum_{i=5}^{10} \partial_{\mu}\Psi^{i}\partial^{\mu}\Psi^{i}
+\frac{1}{4} \sum_{i=1}^{4} \partial_{\mu}\Psi^{i}\partial^{\mu}\Psi^{i}
-\frac{1}{8} \sum_{i=1}^{4} \partial_{\mu}\Psi^{i} \sum_{j=1}^{4} \partial^{\mu}\Psi^{j}\nonumber\\
 &\equiv&
\frac{1}{2} \sum_{i,j} \partial_{\mu}\Psi^{i}V_{ij}\partial^{\mu}\Psi^{j},
\label{quadratic}
\end{eqnarray}
where the symmetric matrix 
\begin{equation}
V_{ij} =
\left\lbrace \begin{array}{ccc}
\delta_{ij} & \mathrm{for} & i,j \geq 5\\
\left(\begin{array}{cccc}
1/4&-1/4&-1/4&-1/4\\-1/4&1/4&-1/4&-1/4\\-1/4&-1/4&1/4&-1/4\\-1/4&-1/4&-1/4&1/4\\
\end{array}\right) & \mathrm{for} & 1 \leq i,j \leq 4
\end{array} \right.
\end{equation}
The propagator itself is computed by Fourier transforming to momentum space, which as usual transforms (\ref{quadratic}) into an algebraic equation in the momentum $k^{\mu}$. The propagator then solves the (formal) matrix equation $k^{2}\mathbf{VP} = \mathbf{I}$ where the identity matrix is now symmetrised as above {\it i.e.} $\mathbf{P} = \frac{1}{k^{2}}\mathbf{V}^{-1}$.
Inverting $\mathbf{V}$ is simple enough and gives,
\begin{equation}
(V^{-1})_{ij} =
\left\lbrace \begin{array}{ccc}
\delta_{ij} & if & i,j \geq 5\\
\left(\begin{array}{cccc}
1&-1&-1&-1\\-1&1&-1&-1\\-1&-1&1&-1\\-1&-1&-1&1\\
\end{array}\right) & if & 1 \leq i,j \leq 4
\end{array} \right.
\end{equation}
Then translating back to tensor indices with the same key, \eqref{Propagator_Key}, we find that $(V^{-1})_{ij}$ corresponds to the matrix 
\begin{equation}
\eta^{\mu\alpha}\eta^{\nu\beta}+\eta^{\mu\beta}\eta^{\nu\alpha}-\eta^{\mu\nu}\eta^{\alpha\beta},
\end{equation}
which, in turn gives the celebrated graviton propagator in GR,
\begin{equation} \label{Propagator_GR}
P^{\mu_{1}\nu_{1},\mu_{2}\nu_{2}}(k) = \frac{\eta^{\mu_{1}\mu_{2}}\eta^{\nu_{1}\nu_{2}}+\eta^{\mu_{1}\nu_{2}}\eta^{\nu_{1}\mu_{2}}-\eta^{\mu_{1}\nu_{1}}\eta^{\mu_{2}\nu_{2}}}{k^{2}}.
\end{equation}
Using the same translation between tensor and vector indices as above, the propagator for unimodular gravity can be derived in a similar way. The gauge fixed expression for the quadratic Lagrangian in \eqref{Lagrangian_Perturbative_TFG_dD} reads
\begin{eqnarray}
  \hat{\mathcal{L}}_{2} &=& \frac{1}{2} \partial_{\gamma} h_{\alpha\beta}\, \hat{V}^{\alpha\beta\mu\nu}\, 
  \partial^{\gamma} h_{\mu\nu},
\end{eqnarray}
with $\displaystyle \hat{V}^{\alpha\beta\mu\nu} \equiv \frac{3}{16} \eta^{\alpha\beta}\eta^{\mu\nu} - \frac{1}{2} \eta^{\alpha\mu}\eta^{\beta\nu}$. Again, this can be put into the form,
\begin{eqnarray}
  \hat{\mathcal{L}}_{2}=
  \frac{1}{2} \sum_{i,j} \partial_{\mu}\Psi^{i}\hat{V}_{ij}\partial^{\mu}\Psi^{j},
  \label{quadratic_UG}
\end{eqnarray}
where the symmetric matrix 
\begin{equation}
\hat{V}_{ij} =
\left\lbrace \begin{array}{ccc}
\delta_{ij} & \mathrm{for} & i,j \geq 5\\
\left(\begin{array}{cccc}
5/16&-3/16&-3/16&-3/16\\-3/16&5/16&-3/16&-3/16\\-3/16&-3/16&5/16&-3/16\\-3/16&-3/16&-3/16&5/16\\
\end{array}\right) & \mathrm{for} & 1 \leq i,j \leq 4
\end{array} \right.
\end{equation}
Inverting $\hat{\mathbf{V}}$ gives,
\begin{equation}
(\hat{V}^{-1})_{ij} =
\left\lbrace \begin{array}{ccc}
\delta_{ij} & if & i,j \geq 5\\
\left(\begin{array}{cccc}
1/2&-3/2&-3/2&-3/2\\-3/2&1/2&-3/2&-3/2\\-3/2&-3/2&1/2&-3/2\\-3/2&-3/2&-3/2&1/2\\
\end{array}\right) & if & 1 \leq i,j \leq 4
\end{array} \right.
\end{equation}
and, translating back to rank-2 indices with \eqref{Propagator_Key}, we find that $(\hat{V}^{-1})_{ij}$ corresponds to the matrix
\begin{equation}
\eta^{\mu\alpha}\eta^{\nu\beta}+\eta^{\mu\beta}\eta^{\nu\alpha}-\frac{3}{2}\eta^{\mu\nu}\eta^{\alpha\beta},
\end{equation}
which leads to the unimodular gravity propagator
\begin{equation} 
  \label{Propagator_TFG}
  \hat{P}^{\mu_{1}\nu_{1},\mu_{2}\nu_{2}}(k) = \frac{\eta^{\mu_{1}\mu_{2}}\eta^{\nu_{1}\nu_{2}}+  
  \eta^{\mu_{1}\nu_{2}}\eta^{\nu_{1}\mu_{2}}-\frac{3}{2}\eta^{\mu_{1}\nu_{1}}\eta^{\mu_{2}\nu_{2}}}{k^{2}}.
\end{equation}
As alluded to earlier, the differences between the quadratic actions of GR and UG are such that the index structure of the propagators are the same with the only change coming in one of the coefficients in the numerator of \eqref{Propagator_GR}.

\subsection{Rules} \label{Rules_Feynman}
Having now derived the perturbative Lagrangians for both GR and UG, found the propagators and set up the formalism to calculate the amplitudes it remains only to extract the Feynman rules for graviton scattering. We begin by assigning a particle number to each of the gravitons in a given expression {\it from left to right}. We also designate the left index of a particular graviton by $\mu_{i}$ and its right index by $\nu_{i}$. Then, contracting (left) indices on two particles, say $i$ and $j$, produces a factor $\eta^{\mu_{i}\mu_{j}}$ with similar factors for left-right, right-left and right-right contractions. Similarly, contracting a derivative of graviton $i$ with another graviton, $j$, produces a factor $k_{i}^{\mu_{j}}$ while contracting indices on two derivatives gives a product of the momenta of the corresponding gravitons, $k_{i}\cdot k_{j}$.

As an example, applying the Feynman rules above to the following term encountered at cubic order in the perturbative 
GR Lagrangian,
\begin{eqnarray}
  h^{\alpha\gamma} \partial_{\gamma} h_{\beta\delta} \partial^{\delta} h_{\alpha}^{\beta},
\end{eqnarray}
results in a cubic vertex factor
\begin{eqnarray}
  k_{2}^{\mu_{1}} k_{3}^{\nu_{2}} \eta^{\mu_{2}\nu_{3}} \eta^{\mu_{1}\mu_{3}}.
\end{eqnarray}
But we then permute the vertex rule through all the permutations of the external legs of the diagram, {\it i.e.} permute $(k_{i},\mu_{i}, \nu_{i})$ through $i=1,2,3$, keeping in mind the symmetry in $(\mu_{i}\nu_{i})$. This particular term has six distinct permutations. To account for this, we introduce the notation $P_{k}$, where $P$ permutes the particle labels among the external legs and $k$ designates the number of distinct permutations. The complete rule for this example then reads
\begin{eqnarray}
  P_{6}\left( k_{2}^{\mu_{1}} k_{3}^{\nu_{2}} \eta^{\mu_{2}\nu_{3}} \eta^{\mu_{1}\mu_{3}}\right).
\end{eqnarray}
We are now ready to compute the scattering amplitudes for both GR and UG.

\subsection{Three Point Amplitude}
We start with the three point amplitudes for both GR and TFG, beginning with the
graviton 3-vertex given by the Feynman diagram in Figure 1.
\begin{figure}[htb!]
\begin{center}
\includegraphics[width = 4.5cm,height = 4.0cm]{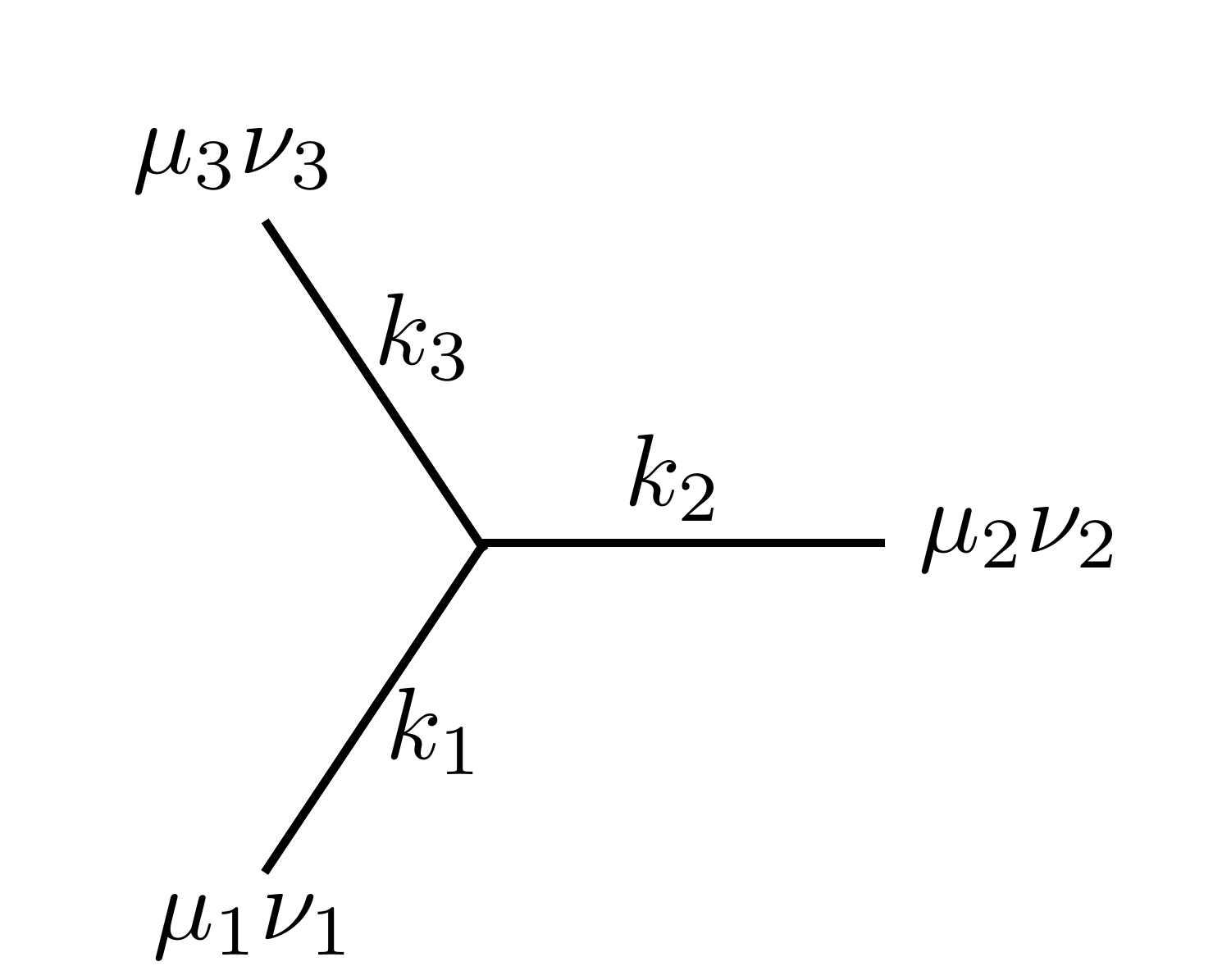}
\caption{3-graviton vertex}
\end{center}
\end{figure}
Given the gauge fixed cubic Lagrangian $\mathcal{L}_{3}$ in GR, \eqref{Lagrangian_Perturbative_GR_dD}, 
we extract the following 3-vertex rule
\begin{eqnarray}
\label{Vertex_3_GR}
V^{\mu_{1}\nu_{1};\mu_{2}\nu_{2};\mu_{3}\nu_{3}}(k_{1},k_{2},k_{3}) &=&
\left( \frac{1}{4}  P_{6}\left( k_{1}\cdot k_{2} \eta^{\mu_{1}\nu_{1}}\eta^{\mu_{2}\mu_{3}}\eta^{\nu_{2}\nu_{3}} \right)\right.
- P_{3}\left( k_{1}\cdot k_{2} \eta^{\nu_{1}\mu_{2}}\eta^{\nu_{2}\mu_{3}}\eta^{\nu_{3}\mu_{1}} \right)\nonumber\\ 
&+& \frac{1}{4} P_{3}\left( k_{1}\cdot k_{2} \eta^{\mu_{1}\mu_{2}}\eta^{\nu_{1}\nu_{2}}\eta^{\mu_{3}\nu_{3}} \right)
- \frac{1}{8} P_{3}\left( k_{1}\cdot k_{2} \eta^{\mu_{1}\nu_{1}}\eta^{\mu_{2}\nu_{2}}\eta^{\mu_{3}\nu_{3}} \right)\nonumber\\
&-& \frac{1}{2} P_{3}\left( k_{1}^{\mu_{3}}k_{2}^{\nu_{3}} \eta^{\mu_{1}\mu_{2}}\eta^{\nu_{1}\nu_{2}} \right)
+\left. \frac{1}{2} P_{6}\left( k_{1}^{\mu_{3}}k_{2}^{\mu_{1}} \eta^{\nu_{1}\mu_{2}}\eta^{\nu_{2}\nu_{3}} \right)\right).
\end{eqnarray}
With this in place, we can now calculate the amplitude by applying the same method as for the gluon 3-amplitude in the appendix. Before deciding on the helicities of the particles it will be useful to first consider the special kinematics of the particles. Depending on the helicity structure, we will either choose $\ket{i}\propto \ket{j}$, which implies $\braket{i}{j}=0$, or $\kets{i}\propto \kets{j}$, which implies $\brakets{i}{j}=0$, for all particles $i$ and $j$. Irrespective of our choice though, terms containing a dot product of momenta $k_{i}\cdot k_{j}$ will vanish since $k_{i}\cdot k_{j}=\grakets{i}{\gamma^{\mu}}{i}\grakets{j}{\gamma_{\mu}}{j}=\frac{1}{2}\braket{i}{j}\brakets{i}{j}$. This allows us to simplify the 3-vertex rule to
\begin{eqnarray}
V^{\mu_{1}\nu_{1};\mu_{2}\nu_{2};\mu_{3}\nu_{3}}(k_{1},k_{2},k_{3}) &=&
- \frac{1}{2} P_{3}\left( k_{1}^{\mu_{3}}k_{2}^{\nu_{3}} \eta^{\mu_{1}\mu_{2}}\eta^{\nu_{1}\nu_{2}} \right)
+ \frac{1}{2} P_{6}\left( k_{1}^{\mu_{3}}k_{2}^{\mu_{1}} \eta^{\nu_{1}\mu_{2}}\eta^{\nu_{2}\nu_{3}} \right).\nonumber
\end{eqnarray}
To illustrate how this computation works in some detail, let's choose a specific case, say $M_{3}(k_{1}^{-},k_{2}^{-},k_{3}^{+})$. For this amplitude, we choose the special kinematics $\kets{i}\propto\kets{j}$ for all particles $i$ and $j$ so that,
\begin{eqnarray}
k_{i}^{\mu_{3}}=\epsilon^{\mu_{3}}k_{i}=-\frac{\grakets{q_{3}}{\gamma^{\mu}}{3}\grakets{i}{\gamma_{\mu}}{i}}{2\sqrt{2}\braket{q_{3}}{3}} = -\frac{\braket{q_{3}}{i}\brakets{3}{i}}{\sqrt{2}\braket{q_{3}}{3}},
\end{eqnarray}
and any term containing $k_{i}^{\mu_{3}}$ or $k_{i}^{\nu_{3}}$ for $i=1,2$ will necessarily vanish, further reducing the 3-vertex rule to
\begin{eqnarray}
\label{Vertex_3_GR_Reduced}
V^{\mu_{1}\nu_{1};\mu_{2}\nu_{2};\mu_{3}\nu_{3}}(k_{1}^{-},k_{2}^{-},k_{3}^{+}) &=&
- \frac{1}{2} \left( k_{2}^{\mu_{1}}k_{3}^{\nu_{1}} \eta^{\mu_{2}\mu_{3}}\eta^{\nu_{2}\nu_{3}} + 
k_{3}^{\mu_{2}}k_{1}^{\nu_{2}} \eta^{\mu_{3}\mu_{1}}\eta^{\nu_{3}\nu_{1}} \right)\nonumber\\
&+& \frac{1}{2} \left( k_{2}^{\mu_{1}}k_{3}^{\mu_{2}} \eta^{\nu_{2}\mu_{3}}\eta^{\nu_{3}\nu_{1}} + 
 k_{1}^{\mu_{2}}k_{3}^{\mu_{1}} \eta^{\nu_{1}\mu_{3}}\eta^{\nu_{3}\nu_{2}} \right).\,\,\,
\end{eqnarray}
Now we rewrite the vertex rule in terms of spinor brackets using the conventions set out in the appendix \ref{Spin-1_Fields}, starting with the decomposition into polarisation vectors so that, for example, $p_{1}^{\mu_{2}} = \epsilon_{\mu}(p_{2}) p^{\mu}_{1}$ and $\eta^{\mu_{1}\mu_{2}} = \epsilon^{\mu}(p_{1}) \epsilon_{\mu}(p_{2})$, followed by the translation into spinor-helicity variables through,
\begin{eqnarray}
  p^{\mu}_{i}
  &=&
  \frac{1}{2}\grakets{1}{\gamma_{\mu}}{1},\nonumber \\
  \epsilon_{\mu}^{+}(p_{1})
  &=&
  -\frac{\grakets{q_{1}}{\gamma_{\mu}}{1}}{\sqrt{2}\braket{q_{1}}{1}}, \\
  \epsilon_{\mu}^{-}(p_{1})
  &=&
  -\frac{\grakets{1}{\gamma_{\mu}}{q_{1}}}{\sqrt{2}\brakets{q_{1}}{1}}, \nonumber
\end{eqnarray}
where here, and in what follows below, the $q_{i}$ are arbitrary reference spinors which will not feature in the final expression for the amplitude. After this initial substitution we can then contract the associated angle and square brackets to give the amplitude in square- and angle-spinor brackets, using the relation
\begin{eqnarray}
  \grakets{i}{\gamma_{\mu}}{j}\grakets{k}{\gamma^{\mu}}{l}
  &=&
  2 \braket{i}{k}\brakets{j}{l}. \nonumber
\end{eqnarray}
This leads to the following expression for the three-point amplitude,
\begin{eqnarray}
  M_{3}(k_{1}^{-},k_{2}^{-},k_{3}^{+})
  &=&
  \frac{-1}{\brakets{q_1}{1}^{2}\brakets{q_2}{2}^{2}\braket{q_3}{3}^{2}}
  \Bigl(
   \braket{2}{1}\brakets{q_2}{1} \braket{2}{3}\brakets{q_2}{3} \braket{1}{q_3}^2\brakets{q_1}{3}^2\nonumber \\
  &+&\braket{1}{2}\brakets{q_1}{2} \braket{1}{3}\brakets{q_1}{3} \braket{2}{q_3}^2\brakets{q_2}{3}^2 
  -\braket{2}{1}\brakets{q_2}{1} \braket{1}{3}\brakets{q_1}{3} \braket{1}{q_3}\brakets{q_1}{3} \braket{2}{q_3}\brakets{q_2}{3}\nonumber\\
  &-&\braket{1}{2}\brakets{q_1}{2} \braket{2}{3}\brakets{q_2}{3} \braket{1}{q_3}\brakets{q_1}{3} \braket{2}{q_3}\brakets{q_2}{3}
  \Bigr). 
\end{eqnarray}
This expression can then be simplified using the antisymmetry property of the spinor brackets along with conservation of momentum which, in spinor-helicity language reads $\sum_{j} \braket{i}{j}\brakets{j}{k} =0$. This gives,
\begin{eqnarray}
  M_{3}(k_{1}^{-},k_{2}^{-},k_{3}^{+}) 
  &=&
  \frac{\braket{1}{2}^{4}}{\braket{q_{3}}{3}^{2}} \left(
  \left(\frac{\braket{q_{3}}{1}}{\braket{1}{3}}\right)^2
  -2\frac{\braket{q_{3}}{1}\braket{q_{3}}{2}}{\braket{1}{3}\braket{2}{3}}
  +\left( \frac{\braket{q_{3}}{2}}{\braket{2}{3}} \right)^{2}
  \right)
\end{eqnarray}
This can then be factorized and the Schouten identity, $\braket{i}{j}\braket{k}{l}+\braket{i}{k}\braket{l}{j}+\braket{i}{l}\braket{j}{k}=0$, can be used to simplify as follows
\begin{eqnarray}
  M_{3}(k_{1}^{-},k_{2}^{-},k_{3}^{+}) 
  &=&
  \frac{\braket{1}{2}^{4}}{\braket{q_{3}}{3}^{2}} 
  \left(
  \frac{\braket{q_{3}}{1}}{\braket{1}{3}}
  -\frac{\braket{q_{3}}{2}}{\braket{2}{3}} 
  \right)^{2}\nonumber\\
  &=&
  \frac{\braket{1}{2}^{4}}{\braket{q_{3}}{3}^{2}} 
  \left(
  \frac{\braket{q_{3}}{1}\braket{2}{3}+\braket{q_{3}}{2}\braket{3}{1}}{\braket{1}{3}\braket{2}{3}}
  \right)^{2}\nonumber\\
  &=&
  \frac{\braket{1}{2}^{4}}{\braket{q_{3}}{3}^{2}} 
  \left(
  \frac{-\braket{q_{3}}{3}\braket{1}{2}}{\braket{1}{3}\braket{2}{3}}
  \right)^{2}\nonumber\\
  &=&
  \frac{\braket{1}{2}^{6}}{\braket{1}{3}^{2}\braket{2}{3}^{2}}. 
\end{eqnarray}
At this point, we note that, first, the $q_{i}$'s have dropped out as promised and, second, the final result depends only on {\it angle} brackets without a square bracket in sight. This is, of course, a consequence of the 3-particle special kinematics. Finally, and more to the point of this article, comparing this to the corresponding one in appendix \ref{Spin-1_Fields} for the color-ordered 3-point gluon scattering amplitude in Yang-Mills theory,
\begin{eqnarray}
  A_{3}\left[1^{-},2^{-},3^{+}\right] = \frac{\braket{1}{2}^{3}}{\braket{2}{3}\braket{3}{1}},
\end{eqnarray}
indeed shows that
\begin{align}
   \label{KLT-3-pt}
   M_{3}(k_{1}, k_{2}, k_{3})=\left( A_{3}[k_{1}, k_{2}, k_{3}] \right)^{2},
\end{align}
which is nothing but the celebrated KLT relation between GR and Yang-Mills theory at 3-points. 
\subsubsection{UG}

Now let's apply these methods to the perturbative unimodular Lagrangian, \eqref{Lagrangian_Perturbative_TFG_dD}. Extracting the 3-vertex rule from the Lagrangian gives in this case,
\begin{eqnarray}
\label{Vertex_3_TFG}
\hat{V}^{\mu_{1}\nu_{1};\mu_{2}\nu_{2};\mu_{3}\nu_{3}}(p_{1},p_{2},p_{3}) &=&
  \frac{1}{8} P_{6}\left( p_{1}\cdot p_{2} \eta^{\mu_{1}\nu_{1}}\eta^{\mu_{2}\mu_{3}}\eta^{\nu_{2}\nu_{3}} \right)
- P_{3}\left( p_{1}\cdot p_{2} \eta^{\nu_{1}\mu_{2}}\eta^{\nu_{2}\mu_{3}}\eta^{\nu_{3}\mu_{1}} \right)\nonumber\\
&+& \frac{3}{8} P_{3}\left( p_{1}\cdot p_{2} \eta^{\mu_{1}\mu_{2}}\eta^{\nu_{1}\nu_{2}}\eta^{\mu_{3}\nu_{3}} \right)
- \frac{5}{64} P_{3}\left( p_{1}\cdot p_{2} \eta^{\mu_{1}\nu_{1}}\eta^{\mu_{2}\nu_{2}}\eta^{\mu_{3}\nu_{3}} \right)
\nonumber\\
&-& \frac{1}{2} P_{3}\left( p_{1}^{\mu_{3}}p_{2}^{\nu_{3}} \eta^{\mu_{1}\mu_{2}}\eta^{\nu_{1}\nu_{2}} \right)
+ \frac{1}{2} P_{6}\left( p_{1}^{\mu_{3}}p_{2}^{\mu_{1}} \eta^{\nu_{1}\mu_{2}}\eta^{\nu_{2}\nu_{3}} \right)
\nonumber\\
&+& \frac{1}{16} P_{3}\left( p_{1}^{\mu_{3}}p_{2}^{\nu_{3}} \eta^{\mu_{1}\nu_{1}}\eta^{\mu_{2}\nu_{2}} \right)
- \frac{1}{8} P_{6}\left( p_{1}^{\mu_{2}}p_{2}^{\mu_{3}} \eta^{\mu_{1}\nu_{1}}\eta^{\nu_{2}\nu_{3}} \right).
\end{eqnarray}
As in the case of GR, the 3-particle special kinematics kills off any term with a momentum dot product, leaving us with
\begin{eqnarray}
\hat{V}^{\mu_{1}\nu_{1};\mu_{2}\nu_{2};\mu_{3}\nu_{3}}(p_{1},p_{2},p_{3}) &=&
- \frac{1}{2} P_{3}\left( p_{1}^{\mu_{3}}p_{2}^{\nu_{3}} \eta^{\mu_{1}\mu_{2}}\eta^{\nu_{1}\nu_{2}} \right)
+ \frac{1}{2} P_{6}\left( p_{1}^{\mu_{3}}p_{2}^{\mu_{1}} \eta^{\nu_{1}\mu_{2}}\eta^{\nu_{2}\nu_{3}} \right) \nonumber \\
&{}&
+ \frac{1}{16} P_{3}\left( p_{1}^{\mu_{3}}p_{2}^{\nu_{3}} \eta^{\mu_{1}\nu_{1}}\eta^{\mu_{2}\nu_{2}} \right)
- \frac{1}{8} P_{6}\left( p_{1}^{\mu_{2}}p_{2}^{\mu_{3}} \eta^{\mu_{1}\nu_{1}}\eta^{\nu_{2}\nu_{3}} \right). \nonumber
\end{eqnarray}
We can now choose the same helicities as in the GR calculation, i.e. $M_{3}(p_{1}^{-},p_{2}^{-},p_{3}^{+})$, which forces the 3-particle special kinematics to be $\kets{i}\propto\kets{j}$ for all particles i and j, eliminating all terms containing $p_{i}^{\mu_{3}}$ or $p_{i}^{\nu_{3}}$ for $i=1,2$ due to the antisymmetry of the square- and angle-spinor brackets. At this point, we deviate from the GR computation, noticing that any trace of the positive helicity particle will also vanish, getting rid of any terms containing $\eta^{\mu_{3}\nu_{3}}$, thereby reducing the 3-vertex rule to
\begin{eqnarray}
\label{Vertex_3_TFG_Reduced}
\hat{V}^{\mu_{1}\nu_{1};\mu_{2}\nu_{2};\mu_{3}\nu_{3}}(p_{1}^{-},p_{2}^{-},p_{3}^{+}) &=
- \frac{1}{2} \left( p_{2}^{\mu_{1}}p_{3}^{\nu_{1}} \eta^{\mu_{2}\mu_{3}}\eta^{\nu_{2}\nu_{3}} 
+ p_{3}^{\mu_{2}}p_{1}^{\nu_{2}} \eta^{\mu_{3}\mu_{1}}\eta^{\nu_{3}\nu_{1}} \right)\\
&
+ \frac{1}{2} \left( p_{2}^{\mu_{1}}p_{3}^{\mu_{2}} \eta^{\nu_{2}\mu_{3}}\eta^{\nu_{3}\nu_{1}} 
+ p_{1}^{\mu_{2}}p_{3}^{\mu_{1}} \eta^{\nu_{1}\mu_{3}}\eta^{\nu_{3}\nu_{2}} \right),
\end{eqnarray}
which is, of course, equivalent to the rule for the 3-vertex in GR. Since we are considering the same external states as in the GR case, we can follow the same substitution rules when converting to spinor variables. This, along with the fact that the vertex expressions are equivalent, yields,
\begin{eqnarray}
  \hat{M}_{3}(p_{1}^{-},p_{2}^{-},p_{3}^{+}) =
  \frac{\braket{1}{2}^{6}}{\braket{1}{3}^{2}\braket{2}{3}^{2}},
\end{eqnarray}
and confirms the KLT relations to 3-points in Unimodular Gravity.
\subsection{Four point Amplitude}
Now let's consider four graviton scattering. As in the previous computation, we will focus on the detailed calculation of the 4-point amplitude in GR, identify the differences with UG and then deduce the associated amplitude in umimodular gravity. We will focus on the maximal helicity violating (MHV) amplitude where all but two of the gravitons have one helicity. At tree level, the complete amplitude receives contributions from four distinct diagrams that can be constructed for the choice of particles. These are the basic four graviton vertex, and the $s$, $t$ and $u$ channel respectively. As usual, we will take all momenta to be outgoing.
To compute the 4-point amplitude, we start with the gauge fixed quartic Lagrangian in the perturbation series\eqref{Lagrangian_Perturbative_GR_dD}. Following the same reasoning as we used in the 3-point amplitude computation, we can extract the four-vertex expression for GR.
\begin{eqnarray}
\label{Vertex_4_GR}
  &{}& V^{\mu_{1}\nu_{1};\mu_{2}\nu_{2};\mu_{3}\nu_{3};\mu_{4}\nu_{4}}(p_{1},p_{2},p_{3},p_{4}) =
  \frac{1}{4} P_{24}(p_{3}\cdot p_{4} \eta^{\mu_1\mu_4}\eta^{\nu_1\mu_3}\eta^{\mu_2\nu_4}\eta^{\nu_2\nu_3})  
  \nonumber\\
  &-& \frac{1}{4} P_{24}(p_{3}\cdot p_{4} \eta^{\mu_1\mu_4}\eta^{\nu_1\nu_4}\eta^{\mu_2\mu_3}\eta^{\nu_2\nu_3})
  - \frac{1}{16} P_{24}(p_{3}\cdot p_{4} \eta^{\mu_1\mu_2}\eta^{\nu_1\nu_2}\eta^{\mu_3\mu_4}\eta^{\nu_3\nu_4})  
  \nonumber\\
  &-& \frac{1}{32} P_{24}(p_{3}\cdot p_{4} \eta^{\mu_1\nu_1}\eta^{\mu_2\nu_2}\eta^{\mu_3\mu_4}\eta^{\nu_3\nu_4})
  - \frac{1}{2} P_{24}(p_{3}^{\mu_1} p_{4}^{\mu_3} \eta^{\nu_1\mu_2}\eta^{\nu_2\mu_4}\eta^{\nu_3\nu_4})\nonumber\\
  &+& \frac{1}{2} P_{24}(p_{3}^{\mu_2} p_{4}^{\mu_3} \eta^{\mu_1\nu_1}\eta^{\nu_2\mu_4}\eta^{\nu_3\nu_4})
  + \frac{1}{2} P_{24}(p_{3}\cdot p_{4} \eta^{\mu_1\mu_2}\eta^{\nu_1\mu_3}\eta^{\nu_2\mu_4}\eta^{\nu_3\nu_4})\nonumber\\
  &-& \frac{1}{4} P_{24}(p_{3}\cdot p_{4} \eta^{\mu_1\nu_1}\eta^{\mu_2\mu_4}\eta^{\nu_2\mu_3}\eta^{\mu_3\nu_4})
  - \frac{1}{2} P_{24}(p_{3}^{\mu_2} p_{4}^{\mu_1} \eta^{\nu_1\mu_3}\eta^{\nu_2\mu_4}\eta^{\nu_3\nu_4})\nonumber\\
 &+& \frac{1}{4} P_{24}(p_{3}^{\mu_1} p_{4}^{\mu_2} \eta^{\nu_1\nu_2}\eta^{\mu_3\mu_4}\eta^{\nu_3\nu_4})
  - \frac{1}{8} P_{24}(p_{3}^{\mu_2} p_{4}^{\nu_2} \eta^{\mu_1\nu_1}\eta^{\mu_3\mu_4}\eta^{\nu_3\nu_4})\nonumber\\
 &-& \frac{1}{8} P_{24}(p_{3}^{\mu_1} p_{4}^{\mu_2} \eta^{\nu_1\nu_2}\eta^{\mu_3\nu_3}\eta^{\mu_4\nu_4})
 + \frac{1}{2} P_{24}(p_{3}^{\mu_4} p_{4}^{\mu_2} \eta^{\mu_1\mu_3}\eta^{\nu_1\nu_3}\eta^{\nu_2\nu_4})\nonumber\\
 &+& \frac{1}{8} P_{24}(p_{3}^{\mu_4} p_{4}^{\mu_3} \eta^{\mu_1\mu_2}\eta^{\nu_1\nu_2}\eta^{\nu_3\nu_4})
 - \frac{1}{16} P_{24}(p_{3}^{\mu_4} p_{4}^{\mu_3} \eta^{\mu_1\nu_1}\eta^{\mu_2\nu_2}\eta^{\nu_3\nu_4})\nonumber\\
 &+& \frac{1}{4} P_{24}(p_{3}^{\mu_4} p_{4}^{\mu_3} \eta^{\mu_1\nu_1}\eta^{\mu_2\nu_3}\eta^{\nu_2\nu_4})
  - \frac{1}{4} P_{24}(p_{3}^{\mu_2} p_{4}^{\nu_2} \eta^{\mu_1\mu_3}\eta^{\nu_1\nu_3}\eta^{\mu_4\nu_4})\nonumber\\
  &+& \frac{1}{4} P_{24}(p_{3}^{\mu_1} p_{4}^{\mu_3} \eta^{\nu_1\mu_2}\eta^{\nu_2\nu_3}\eta^{\mu_4\nu_4})
   - \frac{1}{4} P_{24}(p_{3}^{\mu_2} p_{4}^{\mu_3} \eta^{\mu_1\nu_1}\eta^{\mu_2\nu_3}\eta^{\mu_4\nu_4})
   \nonumber\\
  &+& \frac{1}{4} P_{24}(p_{3}^{\mu_1} p_{4}^{\mu_2} \eta^{\nu_1\mu_3}\eta^{\nu_2\nu_3}\eta^{\mu_4\nu_4})
  + \frac{1}{2} P_{24}(p_{3}^{\mu_2} p_{4}^{\nu_2} \eta^{\mu_1\mu_3}\eta^{\nu_1\mu_4}\eta^{\nu_3\nu_4})\nonumber\\
  &-& \frac{1}{2} P_{24}(p_{3}^{\mu_4} p_{4}^{\mu_2} \eta^{\mu_1\mu_3}\eta^{\nu_1\nu_4}\eta^{\nu_2\nu_3})
   - \frac{1}{4} P_{24}(p_{3}\cdot p_{4} \eta^{\mu_1\mu_2}\eta^{\nu_1\mu_4}\eta^{\nu_2\nu_4}\eta^{\mu_3\nu_3})
   \nonumber\\
  &+& \frac{1}{8} P_{24}(p_{3}\cdot p_{4} \eta^{\mu_1\nu_1}\eta^{\mu_2\mu_4}\eta^{\nu_2\nu_4}\eta^{\mu_3\nu_3}).
\end{eqnarray}
Unlike with the 3-point computation, this is not sufficient since there are also the s-, t- and u-channel diagrams that need to be evaluated. This is, however, easily taken care of with the contraction of two appropriate three point vertex rules. For example, for the s-channel diagram the appropriate vertex factor is given by
\begin{eqnarray}
  V^{\mu_{1}\nu_{1};\mu_{2}\nu_{2};\mu_{s}\nu_{s}}(p_{1},p_{2},p_{s}) V^{\mu_{s}\nu_{s};\mu_{3}\nu_{3};  
  \mu_{4}\nu_{4}}(p_{s},p_{3},p_{4}),
\end{eqnarray}
where the contraction between the two three-vertices is taken over the ``particle" label $s$. Momentum conservation relates its momentum to the external particle momenta through $p_{s} = -p_{1}-p_{2}=p_{3}+p_{4}$. The propagators of the two 3-vertices containing the internal graviton line act together as a place holder for the particle propagator of the theory ultimately sewing together the correct factors of the two three-point vertices. Take, for example, the term
\begin{eqnarray}
  \left( p_{1}^{\mu_{s}}p_{2}^{\nu_{s}} \eta^{\mu_{1}\mu_{2}}\eta^{\nu_{1}\nu_{2}} \right)
  \left( p_{s}^{\mu_{4}}p_{3}^{\nu_{4}} \eta^{\mu_{s}\mu_{3}}\eta^{\nu_{s}\nu_{3}} \right).  
\end{eqnarray}
We first expand this explicitly (including the index structure) as,
\begin{eqnarray}
  \eta^{\mu_{1}\mu_{2}}\eta^{\nu_{1}\nu_{2}}p_{s}^{\mu_{4}}p_{3}^{\nu_{4}}
  \left(p_{1}\right)_{\mu}\left(p_{2}\right)_{\nu}\left(\epsilon^{\mu_{3}}\right)_{\alpha}\left(\epsilon^{\nu_{3}}\right)_{\beta}
  \left(\epsilon^{\mu_{s}}\right)^{\mu}\left(\epsilon^{\nu_{s}}\right)^{\nu}\left(\epsilon^{\mu_{s}}\right)^{\alpha}\left(\epsilon^{\nu_{s}}\right)^{\beta}.
\end{eqnarray}
Then replacing the internal momentum $p_{s}$ with the appropriate representation in the external momenta $p_{1},p_{2},p_{3},p_{4}$, and the factor $\left(\epsilon^{\mu_{s}}\right)^{\mu}\left(\epsilon^{\nu_{s}}\right)^{\nu}\left(\epsilon^{\mu_{s}}\right)^{\alpha}\left(\epsilon^{\nu_{s}}\right)^{\beta}$ with the particle propagator of the theory \eqref{Propagator_GR}, in this case,
\begin{eqnarray}
\left(\epsilon^{\mu_{s}}\right)^{\mu}\left(\epsilon^{\nu_{s}}\right)^{\nu}\left(\epsilon^{\mu_{s}}\right)^{\alpha}\left(\epsilon^{\nu_{s}}\right)^{\beta}
=
P^{\mu\nu\alpha\beta}
=
\frac{\eta^{\mu\alpha}\eta^{\nu\beta}+\eta^{\mu\beta}\eta^{\nu\alpha}-\eta^{\mu\nu}\eta^{\alpha\beta}}{s_{12}},
\end{eqnarray}
where $s_{ij}\equiv -(p_{i}+p_{j})^{2}$. This process is repeated for all other terms in the s-channel amplitude as well as the t- and u-channels, with appropriate choice of momentum in the denominator. 
\begin{figure}[htb!]
\begin{center}
\includegraphics[width = 14.0cm,height = 10.0cm]{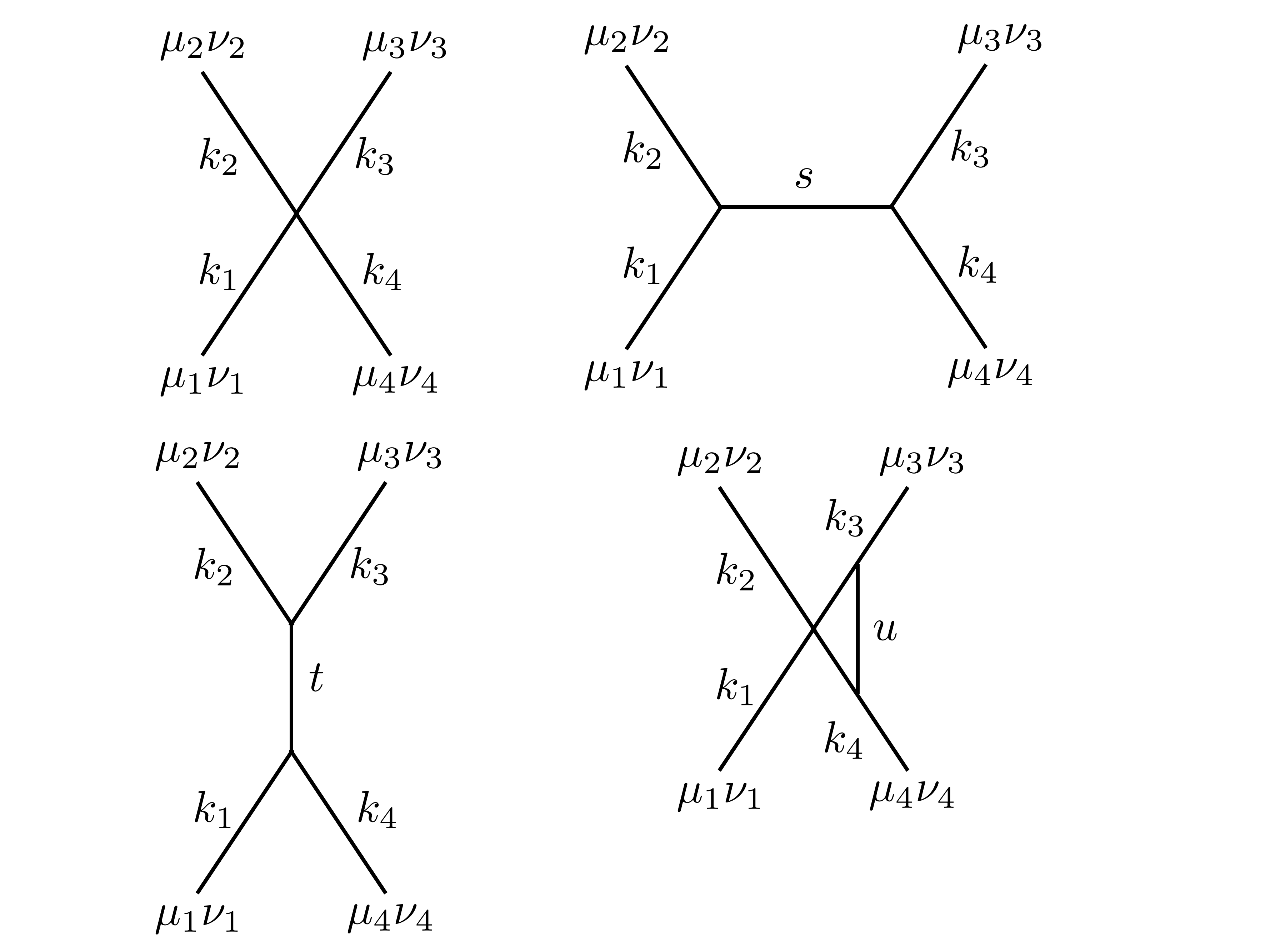}
\caption{4-graviton scattering diagrams}
\end{center}
\end{figure}

Now on to the amplitude calculation. We choose the helicities to be $h_{1}=h_{2}=-2$ and $h_{3}=h_{4}=+2$ ensuring that we have a non-vanishing amplitude since it is MHV. By necessity we assign values to the arbitrary reference momenta for each of the external legs. One such choice, $q_1=q_2=p_4$ and $q_3=q_4=p_1$, will allow us to simplify the expressions we need to calculate to a more manageable size. It is needed to calculate the the factors that will survive this choice of reference momenta. Of all the possible contractions between external particles, the only non-vanishing factors are
\begin{eqnarray}
  \eta^{\mu_{2}\mu_{3}}&=&\frac{\braket{2}{1}\brakets{4}{3}}{\brakets{4}{2}\braket{1}{3}}, \qquad
  p_{2}^{\mu_{1}} = -\frac{\braket{1}{2}\brakets{4}{2}}{\sqrt{2}\brakets{4}{1}}, \qquad
  p_{3}^{\mu_{1}} = -\frac{\braket{1}{3}\brakets{4}{3}}{\sqrt{2}\brakets{4}{1}}, \nonumber\\
  p_{1}^{\mu_{2}} &=& -\frac{\braket{2}{1}\brakets{4}{1}}{\sqrt{2}\brakets{4}{2}}, \qquad
  p_{3}^{\mu_{2}} = -\frac{\braket{2}{3}\brakets{4}{3}}{\sqrt{2}\brakets{4}{2}}, \qquad
  p_{2}^{\mu_{3}} = -\frac{\braket{1}{3}\brakets{3}{2}}{\sqrt{2}\braket{1}{3}}, \\
  p_{4}^{\mu_{3}} &=& -\frac{\braket{1}{4}\brakets{3}{4}}{\sqrt{2}\braket{1}{3}}, \qquad
  p_{2}^{\mu_{4}} = -\frac{\braket{1}{2}\brakets{4}{2}}{\sqrt{2}\braket{1}{4}}, \qquad
  p_{3}^{\mu_{4}} = -\frac{\braket{1}{3}\brakets{4}{3}}{\sqrt{2}\braket{1}{4}}. \nonumber
\end{eqnarray}
Evidently with these choices, the explicit four-point vertex as well as the u-channel diagram both give no contribution to the 4-point amplitude while the remaining two diagrams are greatly simplified giving a final result of
\begin{eqnarray}
M_{4}(p_{1}^{-},p_{2}^{-},p_{3}^{+},p_{4}^{+})
&=&
\frac{\braket{1}{2}^{4} \braket{1}{3}^{2} \brakets{2}{4}^{2} \brakets{3}{4}^{4} \left( \braket{1}{2} \brakets{1}{2} + \braket{1}{3} \brakets{1}{3} \right)}{\braket{1}{4}^{2} \braket{2}{4}^{2} \brakets{1}{3}^{2} \brakets{1}{4}^{2} \braket{1}{2} \brakets{1}{2} \braket{1}{3} \brakets{1}{3} }\nonumber\\
&=&
\frac{\braket{1}{2}^{7} \brakets{1}{2}}{\braket{1}{3} \braket{1}{4} \braket{2}{3} \braket{2}{4} \braket{3}{4}^{2} }.
\end{eqnarray}
On the other hand, the corresponding color-ordered tree-level 4-point MHV gluon scattering amplitude (see for example \cite{Elvang:2013cua}), with the same helicity choice as above, is given by
\begin{eqnarray}
  A_{4}[1^{-},2^{-},3^{+},4^{+}] = \frac{\braket{1}{2}^{4}}{\braket{1}{2}\braket{2}{3}\braket{3}{4}\braket{4}{1}},
\end{eqnarray}
so that
\begin{eqnarray}
  A_{4}[1^{-},2^{-},3^{+},4^{+}] A_{4}[1^{-},2^{-},4^{+},3^{+}] = 
  \frac{\braket{1}{2}^{7} \brakets{1}{2}}{\braket{1}{3} \braket{1}{4} \braket{2}{3} \braket{2}{4} \braket{3}{4}^{2} }\,\,
  \frac{1}{\braket{1}{2}\brakets{1}{2}}.
\end{eqnarray}
In other words, dropping the helicity labels on the scattering particles,
\begin{eqnarray}
  M_{4}^{tree}(1234) = -s_{12}A_{4}^{tree}[1234]\,A_{4}^{tree}[1243],
\end{eqnarray}
precisely as expected for the KLT relations at 4-points. 
\subsubsection{UG}
To compute the four-point amplitude in unimodular gravity we {\it could} follow the same procedure the GR case, noticing that the 4-vertex given by,
\begin{eqnarray}
\label{Vertex_4_UG}
&{}&V^{\mu_{1}\nu_{1};\mu_{2}\nu_{2};\mu_{3}\nu_{3};\mu_{4}\nu_{4}}(p_{1},p_{2},p_{3},p_{4}) =\nonumber\\
&{}&\frac{1}{4} P_{24}(p_{3}\cdot p_{4} \eta^{\mu_1\mu_4}\eta^{\nu_1\mu_3}\eta^{\mu_2\nu_4}\eta^{\nu_2\nu_3})
- \frac{1}{4} P_{24}(p_{3}\cdot p_{4} \eta^{\mu_1\mu_4}\eta^{\nu_1\nu_4}\eta^{\mu_2\mu_3}\eta^{\nu_2\nu_3})\nonumber\\
&-& \frac{1}{16} P_{24}(p_{3}\cdot p_{4} \eta^{\mu_1\mu_2}\eta^{\nu_1\nu_2}\eta^{\mu_3\mu_4}\eta^{\nu_3\nu_4})
+ \frac{7}{64} P_{24}(p_{3}\cdot p_{4} \eta^{\mu_1\nu_1}\eta^{\mu_2\nu_2}\eta^{\mu_3\mu_4}\eta^{\nu_3\nu_4})
\nonumber\\ 
&-& \frac{1}{2} P_{24}(p_{3}^{\mu_1} p_{4}^{\mu_3} \eta^{\nu_1\mu_2}\eta^{\nu_2\mu_4}\eta^{\nu_3\nu_4})
+ \frac{5}{8} P_{24}(p_{3}^{\mu_2} p_{4}^{\mu_3} \eta^{\mu_1\nu_1}\eta^{\nu_2\mu_4}\eta^{\nu_3\nu_4})\nonumber\\
&+& \frac{1}{2} P_{24}(p_{3}\cdot p_{4} \eta^{\mu_1\mu_2}\eta^{\nu_1\mu_3}\eta^{\nu_2\mu_4}\eta^{\nu_3\nu_4})
- \frac{1}{2} P_{24}(p_{3}\cdot p_{4} \eta^{\mu_1\nu_1}\eta^{\mu_2\mu_4}\eta^{\nu_2\mu_3}\eta^{\mu_3\nu_4}) 
\nonumber\\
&-& \frac{1}{2} P_{24}(p_{3}^{\mu_2} p_{4}^{\mu_1} \eta^{\nu_1\mu_3}\eta^{\nu_2\mu_4}\eta^{\nu_3\nu_4})
+ \frac{1}{4} P_{24}(p_{3}^{\mu_1} p_{4}^{\mu_2} \eta^{\nu_1\nu_2}\eta^{\mu_3\mu_4}\eta^{\nu_3\nu_4})
- \frac{1}{4} P_{24}(p_{3}^{\mu_2} p_{4}^{\nu_2} \eta^{\mu_1\nu_1}\eta^{\mu_3\mu_4}\eta^{\nu_3\nu_4})\nonumber\\
&-& \frac{1}{16} P_{24}(p_{3}^{\mu_1} p_{4}^{\mu_2} \eta^{\nu_1\nu_2}\eta^{\mu_3\nu_3}\eta^{\mu_4\nu_4})
+ \frac{1}{2} P_{24}(p_{3}^{\mu_4} p_{4}^{\mu_2} \eta^{\mu_1\mu_3}\eta^{\nu_1\nu_3}\eta^{\nu_2\nu_4})\nonumber\\
&+& \frac{1}{8} P_{24}(p_{3}^{\mu_4} p_{4}^{\mu_3} \eta^{\mu_1\mu_2}\eta^{\nu_1\nu_2}\eta^{\nu_3\nu_4})
- \frac{1}{8} P_{24}(p_{3}^{\mu_4} p_{4}^{\mu_3} \eta^{\mu_1\nu_1}\eta^{\mu_2\nu_2}\eta^{\nu_3\nu_4})\nonumber\\
&+& \frac{1}{8} P_{24}(p_{3}^{\mu_4} p_{4}^{\mu_3} \eta^{\mu_1\nu_1}\eta^{\mu_2\nu_3}\eta^{\nu_2\nu_4})
- \frac{1}{4} P_{24}(p_{3}^{\mu_2} p_{4}^{\nu_2} \eta^{\mu_1\mu_3}\eta^{\nu_1\nu_3}\eta^{\mu_4\nu_4})\nonumber\\
&+& \frac{1}{8} P_{24}(p_{3}^{\mu_1} p_{4}^{\mu_3} \eta^{\nu_1\mu_2}\eta^{\nu_2\nu_3}\eta^{\mu_4\nu_4})
- \frac{1}{4} P_{24}(p_{3}^{\mu_2} p_{4}^{\mu_3} \eta^{\mu_1\nu_1}\eta^{\mu_2\nu_3}\eta^{\mu_4\nu_4})\nonumber\\
&+& \frac{1}{4} P_{24}(p_{3}^{\mu_1} p_{4}^{\mu_2} \eta^{\nu_1\mu_3}\eta^{\nu_2\nu_3}\eta^{\mu_4\nu_4})
+ \frac{1}{2} P_{24}(p_{3}^{\mu_2} p_{4}^{\nu_2} \eta^{\mu_1\mu_3}\eta^{\nu_1\mu_4}\eta^{\nu_3\nu_4})\nonumber\\
&-& \frac{1}{2} P_{24}(p_{3}^{\mu_4} p_{4}^{\mu_2} \eta^{\mu_1\mu_3}\eta^{\nu_1\nu_4}\eta^{\nu_2\nu_3})
- \frac{1}{4} P_{24}(p_{3}\cdot p_{4} \eta^{\mu_1\mu_2}\eta^{\nu_1\mu_4}\eta^{\nu_2\nu_4}\eta^{\mu_3\nu_3})\nonumber\\
&+& \frac{1}{4} P_{24}(p_{3}\cdot p_{4} \eta^{\mu_1\nu_1}\eta^{\mu_2\mu_4}\eta^{\nu_2\nu_4}\eta^{\mu_3\nu_3})
+ \frac{1}{16} P_{24}(p_{3}^{\mu_2} p_{4}^{\nu_2} \eta^{\mu_1\nu_1}\eta^{\mu_3\nu_3}\eta^{\mu_4\nu_4})\nonumber\\
&+& \frac{1}{128} P_{24}(p_{3}\cdot p_{4} \eta^{\mu_1\mu_2}\eta^{\nu_1\nu_2}\eta^{\mu_3\nu_3}\eta^{\mu_4\nu_4})
- \frac{13}{512} P_{24}(p_{3}\cdot p_{4} \eta^{\mu_1\nu_1}\eta^{\mu_2\nu_2}\eta^{\mu_3\nu_3}\eta^{\mu_4\nu_4}),
\end{eqnarray}
is similar to the GR expression but with different constant coefficients {\it etc}. However, recalling that, for the MHV 4-graviton scattering with our choice of reference momenta, only the $s$- and $t$-channel diagrams contributed and these in turn were constructed by sewing together 3-point amplitudes which we've already determined to be the same in UG and GR, we deduce that the 4-point tree-level MHV amplitude in unimodular gravity must be
\begin{eqnarray}
   \hat{M}_{4}^{tree}(p_{1}^{-},p_{2}^{-},p_{3}^{+},p_{4}^{+}) &=&
   \frac{\braket{1}{2}^{7} \brakets{1}{2}}{\braket{1}{3} \braket{1}{4} \braket{2}{3} \braket{2}{4} \braket{3}{4}^{2} }
   \nonumber\\
   &=& -s_{12}A_{4}^{tree}[1234]\,A_{4}^{tree}[1243],
\end{eqnarray}
and the KLT relations hold. It is interesting to note that in the case of gauge theory amplitudes, when considering the color-stripped 4-point amplitude, the diagrams are restricted to those that have no crossing legs, i.e. the $u$-channel diagram is not included. Also, the BCFW recursion relations \cite{Britto:2005fq} allow for the construction of all higher point tree amplitudes from only the 3-vertex and propagators, a property that is known to extend also to GR and that was used to give an explicit proof of the $n$-point KLT relations \cite{BjerrumBohr:2010yc}. With such similarity between the basic building blocks in GR and UG, we anticipate that the same will be true also in unimodular gravity. 

\section{Discussion}
General relativity is an extraordinary theory. In addition to the myriad solutions that describe all manner of gravitational 
phenomena from the precession of the orbit of Mercury to the dynamics of the whole Universe, well known to the gravity 
community, the past several years has seen the application of quantum field theoretic methods expose a completely different 
facet of the theory {\it viz} the underlying mathematical structures that manifest in its scattering amplitudes. One such structure is the KLT relations \eqref{KLT} that express the (tree-level) graviton scattering amplitudes in terms of their gauge theory counterparts. 
In this article, we asked the question: {\it To what extent to these KLT relations extend to `deformations' of GR?} As an example of a case where it does not, let's consider a 3-graviton scattering process in an $f(R) = R^{2}$ gravity theory and compute the MHV amplitude $M_{3}(1^{-}2^{-}3^{+})$. Fortunately, {\it little group scaling} and {\it locality} completely fix the massless 3-particle amplitudes \cite{Elvang:2013cua} as
\begin{eqnarray}
  M_{3}(1^{h_{1}}2^{h_{2}}3^{h_{3}}) = \tilde{\kappa}\braket{1}{2}^{h_{3}-h_{1}-h_{2}}
  \braket{1}{3}^{h_{2}-h_{1}-h_{3}}\braket{2}{3}^{h_{1}-h_{2}-h_{3}},
\end{eqnarray}
where $\tilde{\kappa}$ is the coupling associated with the $R^{2}$ operator and $h_{i} = \pm 2$, the helicities of the gravitons. With $h_{1} = h_{2} = -2$ and $h_{3}=+2$, this gives
\begin{eqnarray}
  M_{3}(1^{-}2^{-}3^{+}) = \tilde{\kappa}\frac{\braket{1}{2}^{6}}{
  \braket{1}{3}^{2}\braket{2}{3}^{2}},
\end{eqnarray}
which looks promising until one realizes that the mass-dimension 2 of the kinematic part requires that the coupling $\tilde{\kappa}$ have mass-dimension -1 in order to ensure that the whole amplitude have the correct mass-dimension of 1. However a quick dimensional analysis check reveals that in this case $[\tilde{\kappa}]=0$. {\it A priori} then, we would not expect generic $f(R)$ gravity theories to exhibit KLT structure. 

Unimodular or Tracefree gravity is different. Since the $\sqrt{|g|}$ does not contribute to the dimensional analysis and can effectively be scaled away, the gravitational coupling associated to UG has the same mass-dimensions as in GR. Consequently, we would indeed expect tree-level results like the KLT relations to hold. Indeed, in this article we have checked explicitly that at least to 5-points it does, even though the structure of the vertex rules in the two theories differ significantly. The overall structure allows the rules to be reduced to the same expressions once a physical set of particle states have been assigned. Consequently, unimodular gravity, the truncation of the Einstein gravity to its tracefree degrees of freedom, also exhibits the same rich structure as GR does, at least with respect to its (tree-level) relationship to gauge theory via the KLT relations. This is to be expected since, classically the two theories are equivalent, and the tree-level amplitudes only encode the semi-classical interactions of the gravitons. 

General relativity and unimodular gravity {\it are} however expected to differ at the quantum level \cite{Alvarez:2012uz}, so the study of graviton scattering is key to breaking the degeneracy. To this end, what is required are the 1-loop and higher amplitudes. This is a formidable task indeed in the context of standard Feynman diagram computations. However, the technological renaissance in amplitude calculations in recent years has seen the development of powerful {\it unitarity methods} (see \cite{Bern:2002kj} and references therein) that use precisely the KLT relations to obtain loop amplitudes from trees. It would be of immense interest to extend our tree-level results to higher loops. Then there is the issue of coupling to matter. One of the key phenomenological motivations for UG is the fact that, unlike in GR, gravity no longer couples to matter potentials \cite{Ellis:2013eqs}. This necessarily means that graviton-matter scattering should differ in the two theories. Again, amplitude technology and the KLT relations in particular allow for such scattering amplitudes to be computed (at least in some restricted cases) \cite{Bern:1999bx}. We would be curious how these amplitudes change in unimodular gravity.

In any event, part of the motivation for this work was to expose these powerful field theoretic methods to a broader community and we hope that, if nothing else, we have succeeded in elucidating further the wonderful legacy left to us by Einstein 100 years ago.

\section{Acknowledgements}
We would like to thank Tim Adamo, Nathan Moynihan and Jean-Philippe Uzan for useful insights and valuable discussion at various stages of this work. We are especially indebited to Nathan Moynihan for bringing exercise 2.34 of \cite{Elvang:2013cua} to our notice. This work is based on the research supported by the South African Research Chairs Initiative of the Department of Science and Technology and National Research Foundation of South Africa as well as the Competitive Programme for Rated Researchers (Grant Number 91552) (AW and DB). DB is also supported by a Masters Bursary from the South African National Institute for Theoretical Physics (NITheP). GFRE is supported by NRF grant 96031. JM is supported in part by the National Research Foundation of South Africa's CPRR program under Grant Number 87667. Opinions, findings and conclusions or recommendations expressed in any publication generated by the NRF supported research is that of the author(s), and that the NRF accepts no liability whatsoever in this regard. 
\appendix
\addcontentsline{toc}{section}{Appendices}
\section*{Appendix}
\section{Spinor Helicity Formalism}\label{Spinor Helicity Formalism}
\renewcommand{\theequation}{\thesection.\arabic{equation}}
\subsection{Formalism}
Our review of the spinor helicity formalism in this appendix will follow quite closely the excellent treatment given in \cite{Elvang:2013cua}. Throughout the paper we use a mostly plus flat metric $\eta_{\mu\nu}=diag(-1,+1,+1,+1)$. Lower case Greek letters designate spacetime indices and run $\mu=\lbrace0,1,2,3\rbrace$ and $a,b,\dot{a},\dot{b}=\lbrace1,2\rbrace$ are 2-spinor indices. We use $\mu_{i}$ and $\nu_{i}$ to respectively label the left- and right-handed spacetime indices of external states of the various amplitudes, where i runs over the number of particles in the interaction. To construct the spinor helicity formalism consider first the Dirac Lagrangian
\begin{eqnarray}
\mathcal{L}_{D} = i \bar{\Psi} \gamma^{\mu} \partial_{\mu} \Psi - m \bar{\Psi} \Psi,  \label{Lagrangian_Dirac}
\end{eqnarray}
where $\Psi$ is a four-spinor and $\bar{\Psi}$ is the Dirac conjugate of $\Psi$ defined by 
\begin{eqnarray} 
\label{Dirac_Conjugate}
\bar{\Psi} = \Psi^{\dagger} \beta,\quad 
\beta = 
\left(\begin{array}{cc}
0& \delta^{\dot{a}}_{\dot{b}} \\
\delta_{a}^{b}& 0
\end{array}\right).
\end{eqnarray}
The Euler-Lagrange equations are, of course, the Dirac equations for $\Psi$ and $\bar{\Psi}$,
\begin{eqnarray} \label{Dirac_Equation}
\left( -i \gamma^{\mu} \partial_{\mu} + m \right) \Psi &=& 0\nonumber\\
\left( i \gamma^{\mu} \partial_{\mu} + m \right) \bar{\Psi} &=& 0.
\end{eqnarray}
These equations admit plane wave solutions which, for $\Psi$ take the form
\begin{eqnarray}
 \Psi \approx u(p) e^{i p x} + v(p) e^{-\i p x}, 
\end{eqnarray}
with $ p^2 = p_{\mu} p^{\mu} = -m^2$ (and similarly for $\bar{\Psi}$). In momentum space the Dirac equations \eqref{Dirac_Equation} reduce to
\begin{eqnarray} \label{Dirac_Equation_Momentum}
\left( \gamma^{\mu} p_{\mu} +m \right) u(p) &=&0\nonumber\\
\left( -\gamma^{\mu} p_{\mu} +m \right) u(p) &=&0.
\end{eqnarray}
Then \eqref{Dirac_Equation_Momentum} has two independent solutions, one for each value 
of $s=\pm$ where, for massless fermions `$\pm$' denotes the particle helicity, i.e.
\begin{eqnarray}
\Psi (x) = \sum_{s=\pm} \int \frac{d^3 p}{(2 \pi)^3 2 E_p} \Bigl( b_s(p) u_s(p) e^{ipx} + d^{\dagger}_s(p) v_s(p) e^{-ipx} \Bigr),
\end{eqnarray}
with a similar expression for $\bar{\Psi}$ involving instead $d_{s}(p)$ and $b^{\dagger}_{s}(p)$. The $d^{(\dagger)}_s$ and $b^{(\dagger)}_d$ are as usual fermionic annihilation and creation operators and $u_s(p)$ and $v_s(p)$ are four component commuting spinors that encode the Grassmann nature of the particles. The vacuum of the theory is defined such that
\begin{eqnarray}
b_{\pm}(p) \ket{0} &=&  0, \nonumber\\
b^{\dagger}_{\pm}(p) \ket{0} &=& \ket{p,\pm}.\nonumber
\end{eqnarray}
For consistency we take all the particles to be outgoing, so that $v_{\pm}(p)$ represents an anti-fermion and $\bar{u}_{\pm} (p)$ a fermion, obtained from the expansion of $\bar{\Psi}$.

In the high-energy limit, in which the rest mass energy of the fermions is negligible relative to their kinetic energy, we can consider the particles as massless. The corresponding massless equations of motion  
\begin{eqnarray} \label{Dirac_Equation_Massless}
  \gamma^{\mu} p_{\mu} v_{\pm}(p)&=&0,\nonumber\\ 
  \bar{u}_{\pm}(p) \gamma^{\mu} p_{\mu} &=&0,
\end{eqnarray}
each have two solutions which can be written in terms of 2-component spinors as
\begin{eqnarray} 
\label{Spinors_4}
v_{+}(p) &=& \left( \begin{array}{cc} \kets{p}_{a}\\ 0 \end{array} \right)\nonumber\\
v_{-}(p) &=& \left( \begin{array}{cc} 0\\ \ket{p}^{\dot{a}} \end{array} \right)\\
\bar{u}_{+}(p) &=& \left( \begin{array}{c} \bras{p}^{a}, 0 \end{array} \right)\nonumber\\
\bar{u}_{-}(p) &=& \left( \begin{array}{c} 0, \bra{p}_{\dot{a}} \end{array} \right).\nonumber
\end{eqnarray}
These angle- and square-brakets, central to the spinor-helicity notation are nothing but the commuting 2-component spinors. The kets here are outgoing anti-fermions and the bras outgoing fermions. Starting with a momentum 4-vector $p^{\mu}=\left(p^{0},p^{i}\right)=\left(E,p^{i}\right)$ with $p^{\mu}p_{\mu}=-m^{2}$ and the Dirac gamma-matrices defined as usual, 
\begin{eqnarray} 
  \label{Gamma_New}
  \gamma^{\mu}p_{\mu} = 
  \left( \begin{array}{cc}
  0 & p_{a\dot{b}} \\ p^{\dot{a}b} & 0
  \end{array} \right),
\end{eqnarray}
where the momentum bi-spinors are defined as
\begin{eqnarray} \label{Momentum_bi-spinors}
  p_{a\dot{b}} = p_{\mu} \left(\sigma^{\mu}\right)_{a\dot{b}},\quad
  p^{\dot{a}b} = p_{\mu} \left(\bar{\sigma}^{\mu}\right)^{\dot{a}b}.
\end{eqnarray}
The two component spinors that we defined in \eqref{Spinors_4} then solve the massless Weyl equations
\begin{eqnarray}
  \label{Weyl_Equations_Massless}
  p^{\dot{a}b} \,\kets{p}_b = 0,\quad
  \bras{p}^{b}\, p_{b \dot{a}} =0,\quad
  p_{a\dot{b}}\, \ket{p}^{\dot{b}} = 0,\quad
  \bra{p}_{\dot{b}}\, p^{\dot{b} a} =0.
\end{eqnarray}

\subsection{Spinor-helicity properties and identities}
Here we take some time to set out some of the properties and conventions of the spinors defined in \eqref{Spinors_4}. Firstly, we note that the 2-spinor indices are raised and lowered using the antisymmetric Levi-Civita tensor,
\begin{eqnarray}\label{Raise_Lower}
\epsilon_{ab}\bras{p}^{b}= \kets{p}_{a},\quad
\epsilon_{\dot{a}\dot{b}}\ket{p}^{\dot{b}} = \bra{p}_{\dot{a}},\quad
\epsilon^{ab}\kets{p}_{b}= \bras{p}^{a},\quad
\epsilon^{\dot{a}\dot{b}}\ket{p}_{\dot{b}}= \bra{p}^{\dot{a}}.
\end{eqnarray}
Next we consider the reality conditions of the of the square- and angle-spinors. The spinor field $\bar{\Psi}$ is the Dirac conjugate of $\Psi$. Applying this conjugation to the momentum space Dirac equations \eqref{Dirac_Equation_Momentum} necessitates that $\bar{v}_{\pm}=\bar{u}_{\mp}$ and $u_{\pm}=v_{\mp}$, if the momentum $p^{\mu}$ is to be real valued. This usually goes by the name of {\it crossing symmetry}. For real momenta then, we have that
\begin{eqnarray} \label{Reality_conditions}
\bras{p}^{a}=\left( \ket{p}^{\dot{a}} \right)^{\ast},\quad
\bra{p}_{\dot{a}} = \left( \kets{p}_{a} \right)^{\ast}.
\end{eqnarray}
The Dirac 4-spinors satisfy a spin completeness relation that for $m=0$ reads
\begin{eqnarray}
  \sum_{s=\pm} u_{s}(p)\bar{u}_{s}(p) =\sum_{s=\pm} v_{s}(p)\bar{v}_{s}(p) 
  = -\gamma^{\mu}p_{\mu}.
\end{eqnarray}
Using the crossing symmetry this can be rewritten in spinor-helicity notation as
\begin{eqnarray}
\label{Spinor_completeness}
  \ket{p}\bras{p}+\kets{p}\bra{p}&=&-\gamma^{\mu}p_{\mu},
\end{eqnarray}
or, in terms of the momentum bi-spinors,
\begin{eqnarray}
   p_{a\dot{b}} &=&-\kets{p}_{a}\bra{p}_{\dot{b}},\nonumber\\ 
   p^{\dot{a}b} &=&-\ket{p}^{\dot{a}}\bras{p}^{b}.
\end{eqnarray}
We now introduce the notation that is the basis for writing amplitudes in the spinor helicity formalism, the angle spinor bracket $\braket{p}{q}$ and the square spinor bracket $\brakets{p}{q}$. For two lightlike vectors, $p^{\mu}$ and $q^{\mu}$ these are defined as
\begin{eqnarray}
\label{Spinor_brackets}
\braket{p}{q}=\bra{p}_{\dot{a}}\ket{q}^{\dot{a}},\quad
\brakets{p}{q}=\bras{p}^{a}\kets{q}_{a},
\end{eqnarray}
with all other combinations vanishing. Since the raising and lowering of the spinor indices are done with the completely antisymmetric tensor these brackets are antisymmetric,
\begin{eqnarray}
  \label{Spinor_brackets_antisymmetric}
  \braket{p}{q}=-\braket{q}{p},\quad
  \brakets{p}{q}=-\brakets{q}{p}.
\end{eqnarray}
Reality of the momenta translates into spinor-helicity language as
\begin{eqnarray}
  \left(\brakets{p}{q}\right)^{\ast} = \braket{q}{p}.
\end{eqnarray}
When working with higher spin states and interactions expressions like $\bar{u}_{-}(p)\gamma^{\mu}v_{+}(q)$ are frequently encountered. Using \eqref{Spinors_4} and the definition of the gamma-matrices as in \eqref{Gamma_New}, we define the angle-square bracket,
\begin{eqnarray}
\grakets{p}{\gamma^{\mu}}{q}
=
\bar{u}_{-}(p)\gamma^{\mu}v_{+}(q)
=
\left( \begin{array}{c} 0, \bra{p}_{\dot{a}} \end{array} \right)
\left( \begin{array}{cc}
0 & p_{a\dot{b}} \\ p^{\dot{a}b} & 0
\end{array} \right)
\left( \begin{array}{cc} \kets{p}_{a}\\ 0 \end{array} \right),
\end{eqnarray}
with a similar expression defining $\graket{p}{\gamma^{\mu}}{q}$ while for {\it same helicity} fermions the product vanishes. These angle-square brackets satisfy
\begin{eqnarray}
  \label{Brackets_Angle-Square}
   \grakets{p}{\gamma^{\mu}}{q} &=& \graket{q}{\gamma^{\mu}}{p}, \nonumber\\
   \left(\grakets{p}{\gamma^{\mu}}{q}\right)^{\ast}&=&\grakets{q}{\gamma^{\mu}}{p},\\
   \grakets{p}{P}{q}&=&P_{\mu}\grakets{p}{\gamma^{\mu}}{q}=
   \bra{p_{\dot{a}}}P^{\dot{a}b}\kets{q_{b}}=\bra{p_{\dot{a}}}\left(- 
   \ket{P^{\dot{a}}}\bras{P^{b}} \right)\kets{q_{b}}=-\braket{p}{P}\brakets{P}{q},\nonumber
\end{eqnarray}
where, in the last line we take $P^{\mu}$ to be a lightlike vector. The {\it Fierz identity} is given by
\begin{eqnarray}
\label{Fierz_identity}
  \grakets{p_1}{\gamma^{\mu}}{p_2}\grakets{p_3}{\gamma_{\mu}}{p_4} = 2 \braket{p_1}  
  {p_3}\brakets{p_2}{p_4}.
\end{eqnarray}
From this it follows quite easily that $k^{\mu}=\frac{1}{2}\grakets{k}{\gamma^{\mu}}{k}$ while two lightlike vectors $p^{\mu}$ and $q^{\mu}$ will satisfy
\begin{eqnarray}
  \left( p+q \right)^{2} = 2p\cdot q = \braket{p}{q}\brakets{p}{q}.
\end{eqnarray}
The next important identity encodes the conservation of momentum, which in spinor notation  becomes
\begin{eqnarray}
  \label{CoM}
  \sum_{i=1}^{n} p^{\,\mu}_{i} =
  \sum_{i=1}^{n}\braket{p}{i}\brakets{i}{k} =0.
\end{eqnarray}
Lastly we have the so-called {\it Schouten identity}. This identity encompasses the rather simple fact that three 2-dimensional vectors, say $\ket{p},\ket{q},\ket{k}$, cannot all be linearly independent. Any one of them must be a linear combination of the other two, $\ket{p}=a\ket{q}+b\ket{k}$. We can then `dot in' $\bra{p},\bra{q},\bra{k}$ as appropriate to determine the constant coefficients $a$ and $b$ giving
\begin{eqnarray}
  \label{Schouten}
  \ket{p}\braket{q}{k}+\ket{q}\braket{k}{p}+\ket{k}\braket{p}{q}=0.
\end{eqnarray}
A similar statement also holds for square-spinors

\subsection{Gauge theory} \label{Spin-1_Fields}
Now let's put the formalism to use to (eventually) compute the tree-level scattering amplitudes in pure Yang-Mills gauge theory. We begin with the Yang-Mills Lagrangian in the {\it Gervais-Neveu} gauge in which
\begin{eqnarray} \label{Lagrangian_YM_GN}
  \mathcal{L}_{YM} = \mathrm{Tr}\left(
  -\frac{1}{2} \partial_{\mu} A_{\nu} \partial^{\mu} A^{\nu} -i g \sqrt{2} \partial^{\mu}   
  A^{\nu} A_{\nu} A_{\mu} + \frac{g^2}{4} A^{\mu} A^{\nu} A_{\nu} A_{\mu}
  \right), 
\end{eqnarray}
and $g$ is, as usual the Yang-Mills coupling. To construct the spinor-helicity representation of the spin-1 particles we `dot-in' the photon polarization vectors. These are constructed from the spinor-helicity variables as
\begin{eqnarray}
  \label{Spin_1}
  \epsilon_{-}^{\mu}(p) &=& 
  - \frac{\grakets{p}{\gamma^{\mu}}{q}}{\sqrt{2}\brakets{p}{q}}\nonumber\\
  \epsilon_{+}^{\mu}(p) &=& 
  - \frac{\grakets{q}{\gamma^{\mu}}{p}}{\sqrt{2}\braket{p}{q}},
\end{eqnarray}
with the massless Weyl equation ensuring that $p_{\mu} \epsilon_{\pm}^{\mu}(p) =0$.
In what follows, it will be useful to put aside the kinematic factors of the color structure in order to better analyse the vertex structure and extract the Feynman rules for the {\it color-ordered} amplitudes. Note that color-ordered amplitudes are computed from diagrams with no lines crossed and a fixed ordering of external lines. The 3-vertex expression for example is then given by
\begin{eqnarray}
  V^{\mu_{1}\mu_{2}\mu_{3}}(p_1,p_2,p_3) 
  = -\sqrt{2} \left( \eta^{\mu_{1}\mu_{2}} p_{1}^{\mu_{3}} + \eta^{\mu_{2}\mu_{3}} p_{2}  
  ^{\mu_{1}} + \eta^{\mu_{3}\mu_{1}} p_{3}^{\mu_{2}} \right),
\end{eqnarray}
where each $\eta$ consists of two spin-1 polarisation vectors. The rules for extracting the amplitude from the vertex are as follows:
\begin{itemize}
 \item To any stand alone momentum $p_{i}$, say, we associate a square-angle bracket
  $p^{\,\mu}_{i} \to \frac{1}{2}\grakets{i}{\gamma^{\mu}}{i}$.
 \item For contracted momenta, for example $p_{1}^{\mu_{2}} \to 
 \epsilon_{\mu}(p_{2}) p^{\mu}_{1}$ and
 \item For each $\eta$ factor, say, $\eta^{\mu_{1}\mu_{2}} \to 
 \epsilon^{\mu}(p_{1}) \epsilon_{\mu}(p_{2})$
\end{itemize}
Given this, the amplitude from the vertex expression can be written down as
\begin{eqnarray}
V^{\mu_{1}\mu_{2}\mu_{3}}(p_1,p_2,p_3) 
= -\sqrt{2} \Bigl( \left(\epsilon^{\mu_{1}}\epsilon^{\mu_{2}}\right) \left(\epsilon^{\mu_{3}}p_{1}\right) + \left(\epsilon^{\mu_{2}}\epsilon^{\mu_{3}}\right) \left(\epsilon^{\mu_{1}}p_{2}\right) + \left(\epsilon^{\mu_{3}}\epsilon^{\mu_{1}}\right) \left(\epsilon^{\mu_{2}}p_{3}\right) \Bigr).\nonumber
\end{eqnarray}
With the notation $p_{1}\rightarrow 1$ etc, the associated {\it color-ordered} amplitude is then
\begin{eqnarray}
A_{3}\left[1,2,3\right] = -\sqrt{2}\Bigl( 
\left(\epsilon^{1}\epsilon^{2}\right) \left(\epsilon^{3}p_{1}\right)
+ \left(\epsilon^{2}\epsilon^{3}\right) \left(\epsilon^{1}p_{2}\right)
+ \left(\epsilon^{3}\epsilon^{1}\right) \left(\epsilon^{2}p_{3}\right) 
\Bigr).
\end{eqnarray}
To take the calculation further we need to first assign helicities to the individual particles. For this example we will choose particles 1 and 2 to have helicity -1  and particle 3 to have +1 helicity. We can then substitute from equations \eqref{Spin_1} to get,
\begin{eqnarray}
A_{3}\left[1^{-},2^{-},3^{+}\right] &=&
\frac{1}{2\brakets{q_{1}}{1}\brakets{q_{2}}{2}\braket{q_{3}}{3}} \Bigl(
\grakets{1}{\gamma^{\mu}}{q_{1}}\grakets{2}{\gamma_{\mu}}{q_{2}}\grakets{q_{3}}{\gamma^{\nu}}{3}\grakets{1}{\gamma_{\nu}}{1}\nonumber\\
&+&\grakets{2}{\gamma^{\mu}}{q_{2}}\grakets{q_{3}}{\gamma_{\mu}}{3}\grakets{1}{\gamma^{\nu}}{q_{1}}\grakets{2}{\gamma_{\nu}}{2}
+\grakets{q_{3}}{\gamma^{\mu}}{3}\grakets{1}{\gamma_{\mu}}{q_{1}}\grakets{2}{\gamma^{\nu}}{q_{2}}\grakets{3}{\gamma_{\nu}}{3}\Bigr)\nonumber\\
&=&
\frac{1}{\brakets{q_{1}}{1}\brakets{q_{2}}{2}\braket{q_{3}}{3}} \Bigl(
\braket{1}{2}\brakets{q_{1}}{q_{2}} \braket{q_{3}}{1}\brakets{3}{1}\nonumber\\
&+&\braket{2}{q_{3}}\brakets{q_{2}}{3} \braket{1}{2}\brakets{q_{1}}{2}
+\braket{q_{3}}{1}\brakets{3}{q_{1}} \braket{2}{3}\brakets{q_{2}}{3} \Bigr).
\end{eqnarray}
We now have to consider 3-particle special kinematics. For now, it will be sufficient to consider the expression
\begin{eqnarray}
\braket{1}{2}\brakets{1}{2} = \left( p_{1} + p_{2} \right)^{2}=p_{3}^{2}=0.
\end{eqnarray}
For this to be true, either the angle spinor bracket or square spinor bracket must vanish. We must either choose $\kets{1}\propto \kets{2}\propto \kets{3}$ or $\ket{1}\propto \ket{2}\propto \ket{3}$. The choice is made by considering the dimension of the expression. In this case, we set $\kets{1}\propto \kets{2}\propto \kets{3}$, killing off the first term so that
\begin{eqnarray}
A_{3}\left[1^{-},2^{-},3^{+}\right] =
\frac{-1}{\brakets{q_{1}}{1}\brakets{q_{2}}{2}\braket{q_{3}}{3}} \Bigl(
 \braket{q_{3}}{2}\brakets{3}{q_{2}} \braket{1}{2}\brakets{2}{q_{1}}
+\braket{q_{3}}{1}\brakets{3}{q_{1}} \braket{2}{3}\brakets{3}{q_{2}} \Bigr).
\end{eqnarray}
This result is still dependant on the arbitrary reference spinors $q_{i}$. This can be eliminated by multiplying each term by the appropriate representation of $1$. This allows the use of conservation of momentum to change the structure of the brackets, \eqref{CoM} so that, for example, 
\begin{eqnarray}
\braket{1}{3}\brakets{3}{q_{2}} =
-\braket{1}{2}\brakets{2}{q_{2}}-\braket{1}{1}\brakets{1}{q_{2}} =
-\braket{1}{2}\brakets{2}{q_{2}}.
\end{eqnarray}
Substituting back into the amplitude,
\begin{eqnarray}
  A_{3}\left[1^{-},2^{-},3^{+}\right] &=&
  \frac{-1}{\brakets{q_{1}}{1}\brakets{q_{2}}{2}\braket{q_{3}}{3}} \left(
  \frac{\braket{q_{3}}{2}\braket{1}{2}\brakets{2}{q_{2}} \braket{1}{2}\braket{3}{1}\brakets{1}  
  {q_{1}}}{\braket{1}{3} \braket{3}{2}}
  +\frac{\braket{q_{3}}{1}\braket{2}{1}\brakets{1}{q_{1}} \braket{2}{3}\braket{1}{2}\brakets{2}  
  {q_{2}}}{\braket{2}{3} \braket{1}{3}} \right)\nonumber\\
  &=&
  \frac{-\braket{1}{2}^{2}}{\braket{q_{3}}{3}} \left(
  \frac{\braket{q_{3}}{2}\braket{1}{3} -\braket{q_{3}}{1}\braket{2}{3}}{\braket{1}  
  {3}\braket{2}{3}}\right).
\end{eqnarray}
Finally we apply the Schouten identity \eqref{Schouten} and simplify to get
\begin{eqnarray}
  A_{3}\left[1^{-},2^{-},3^{+}\right] &=&
  \frac{-\braket{1}{2}^{2}}{\braket{q_{3}}{3}} \Bigl(
  \frac{\braket{q_{3}}{2}\braket{1}{3} +\braket{q_{3}}{1}\braket{3}{2}}{\braket{1}  
  {3}\braket{2}{3}} \Bigr)\nonumber\\
  &=&
  \frac{-\braket{1}{2}^{2}}{\braket{q_{3}}{3}} \left(
  \frac{\braket{q_{3}}{3}\braket{1}{2}}{\braket{1}{3}\braket{2}{3}} \right)\nonumber\\
  &=&
  \frac{\braket{1}{2}^{3}}{\braket{2}{3}\braket{3}{1}}.
\end{eqnarray}
This then is the color-ordered 3-point amplitude for non-abelian Yang-Mills gauge theory. This can now be dressed with color factors and interaction strength to obtain the full three-gluon amplitude.


\begin{thebibliography}{99}

\bibitem{Deser:2009fq} 
  S.~Deser,
  ``Gravity from self-interaction redux,''
  Gen.\ Rel.\ Grav.\  {\bf 42}, 641 (2010)
  [arXiv:0910.2975 [gr-qc]].

\bibitem{Callan:1985ia} 
  C.~G.~Callan, Jr., E.~J.~Martinec, M.~J.~Perry and D.~Friedan,
  ``Strings in Background Fields,''
  Nucl.\ Phys.\ B {\bf 262}, 593 (1985).

\bibitem{Kawai:1985xq} 
  H.~Kawai, D.~C.~Lewellen and S.~H.~H.~Tye,
  ``A Relation Between Tree Amplitudes of Closed and Open Strings,''
  Nucl.\ Phys.\ B {\bf 269}, 1 (1986).

\bibitem{Bern:2002kj} 
  Z.~Bern,
  ``Perturbative quantum gravity and its relation to gauge theory,''
  Living Rev.\ Rel.\  {\bf 5}, 5 (2002)
  [gr-qc/0206071].
      
\bibitem{Elvang:2013cua} 
  H.~Elvang and Y.~t.~Huang,
  ``Scattering Amplitudes,''
  arXiv:1308.1697 [hep-th].

\bibitem{Witten:2003nn} 
  E.~Witten,
  ``Perturbative gauge theory as a string theory in twistor space,''
  Commun.\ Math.\ Phys.\  {\bf 252}, 189 (2004)
  [hep-th/0312171].

\bibitem{Arkani-Hamed:2013jha} 
  N.~Arkani-Hamed and J.~Trnka,
  ``The Amplituhedron,''
  JHEP {\bf 1410}, 030 (2014)
  [arXiv:1312.2007 [hep-th]].
  
\bibitem{Weinberg:1988cp} 
  S.~Weinberg,
  ``The Cosmological Constant Problem,''
  Rev.\ Mod.\ Phys.\  {\bf 61}, 1 (1989).
  
\bibitem{Ellis:2010uc} 
  G.~F.~R.~Ellis, H.~van Elst, J.~Murugan and J.~P.~Uzan,
  ``On the Trace-Free Einstein Equations as a Viable Alternative to General Relativity,''
  Class.\ Quant.\ Grav.\  {\bf 28}, 225007 (2011)
  [arXiv:1008.1196 [gr-qc]].

\bibitem{Alvarez:2012uz} 
  E.~Alvarez,
  ``The Weight of matter,''
  JCAP {\bf 1207}, 002 (2012)
  [arXiv:1204.6162 [hep-th]].

\bibitem{Alvarez:2015pla} 
  E.~çlvarez, S.~Gonz‡lez-Mart'n, M.~Herrero-Valea and C.~P.~Mart'n,
  ``Unimodular Gravity Redux,''
  Phys.\ Rev.\ D {\bf 92}, no. 6, 061502 (2015)
  [arXiv:1505.00022 [hep-th]].

\bibitem{Padilla:2014yea} 
  A.~Padilla and I.~D.~Saltas,
  ``A note on classical and quantum unimodular gravity,''
  arXiv:1409.3573 [gr-qc].

\bibitem{LopezVillarejo:2010iq} 
  J.~J.~Lopez-Villarejo,
  ``TransverseDiff gravity is to scalar-tensor as unimodular gravity is to General Relativity,''
  JCAP {\bf 1111}, 002 (2011)
  [arXiv:1009.1023 [hep-th]].

\bibitem{Henneaux:1989zc} 
  M.~Henneaux and C.~Teitelboim,
  ``The Cosmological Constant and General Covariance,''
  Phys.\ Lett.\ B {\bf 222}, 195 (1989).
  
\bibitem{Britto:2005fq} 
  R.~Britto, F.~Cachazo, B.~Feng and E.~Witten,
  ``Direct proof of tree-level recursion relation in Yang-Mills theory,''
  Phys.\ Rev.\ Lett.\  {\bf 94}, 181602 (2005)
  [hep-th/0501052].

\bibitem{BjerrumBohr:2010yc} 
  N.~E.~J.~Bjerrum-Bohr, P.~H.~Damgaard, B.~Feng and T.~Sondergaard,
  ``Proof of Gravity and Yang-Mills Amplitude Relations,''
  JHEP {\bf 1009}, 067 (2010)
  [arXiv:1007.3111 [hep-th]].
  
\bibitem{Ellis:2013eqs} 
  G.~F.~R.~Ellis,
  ``The Trace-Free Einstein Equations and inflation,''
  Gen.\ Rel.\ Grav.\  {\bf 46}, 1619 (2014)
  [arXiv:1306.3021 [gr-qc]].
  
\bibitem{Bern:1999bx} 
  Z.~Bern, A.~De Freitas and H.~L.~Wong,
  ``On the coupling of gravitons to matter,''
  Phys.\ Rev.\ Lett.\  {\bf 84}, 3531 (2000)
  [hep-th/9912033].
   
\end{thebibliography}
\end{document}